\documentclass[a4paper,10pt,twoside]{cpc-hepnp}

\usepackage{multicol}
\usepackage{graphicx}
\usepackage{booktabs}
\usepackage{amssymb,bm,mathrsfs,bbm,amscd}
\usepackage[tbtags]{amsmath}
\usepackage{lastpage}

\usepackage{subfig}
\usepackage{caption}
\usepackage{floatrow}

\begin{document}

\fancyhead[c]{\small Chinese Physics C~~~Vol. 37, No. 1 (2013)
010201} \fancyfoot[C]{\small 010201-\thepage}

\footnotetext[0]{Received 14 March 2009}

\title{A simulation study of a dual-plate in-room PET system for dose verification in carbon ion therapy\thanks{National Natural Science Foundation of China (11205222), West Light Foundation of the Chinese Academy of Sciences (210340XBO), National Major Scientific Instruments and Equipment Development Projects (2011YQ12009604), Youth Innovation Promotion Association, CAS (201330YQO)}}

\author{%
      Chen Ze$^{1,2}$
\quad Hu Zheng-Guo$^{2;1)}$\email{huzg@impcas.ac.cn}%
\quad Xiao Guo-Qing$^{2}$\\
\quad Chen Jin-Da$^{2}$
\quad Zhang Xiu-Ling$^{2}$
\quad Guo Zhong-Yan$^{2}$\\
\quad Sun Zhi-Yu$^{2}$
\quad Huang Wen-Xue$^{2}$
\quad Wang Jian-Song$^{2}$
}
\maketitle

\address{
$^1$ University of Chinese Academy of Sciences, Beijing 100049, China\\
$^2$ Institute of Modern Physics, Chinese Academy of Sciences, Lanzhou 730000, China\\
}

\begin{abstract}
Carbon ion therapy have the ability to overcome the limitation of convertional radiotherapy due to its most energy deposition in selective depth, usually called Bragg peak, which results in increased biological effectiness.
During carbon ion therapy, lots positron emitters such as $^{11}$C, $^{15}$O, $^{10}$C are generated in irradiated tissues by nuclear reactions.
Immediately after patient irradiation, PET scanners can be used to measure the spatial distribution of positron emitters, which can track the carbon beam to the tissue.
In this study, we designed and evaluated an dual-plate in-room PET scanner to monitor patient dose in carbon ion therapy, which is based on GATE simulation platform.
A dual-plate PET is designed to avoid interference with the carbon beam line and with patient positioning.
Its performance was compared with that of four-head and full-ring PET scanners.
The dual-plate, four-head and full-ring PET scanners consisted of 30, 60, 60 detector modules, respectively, with a 36 cm distance between directly opposite detector modules for dose deposition measurements.
Each detector module was consisted of a 24$\times$24 array of 2$\times$2$\times$18 mm$^{3}$ LYSO pixels coupled to a Hamamatsu H8500 PMT.
To esitmate the production yield of positron emitters, a 10$\times$15$\times$15 cm$^{3}$ cuboid PMMA phantom was irradiated with 172, 200, 250 AMeV $^{12}$C beams.
3D images of the activity distribution of the three type scanners are produced by an iterative reconstruction algorithm.
By comparing the longitudinal profile of positron emitters, measured along the carbon beam path, we concluded that the development of a dual-plate PET scanner is feasible to monitor the dose distribution for carbon ion therapy.

\end{abstract}



\footnotetext[0]{\hspace*{-3mm}\raisebox{0.3ex}{$\scriptstyle\copyright$}2013
Chinese Physical Society and the Institute of High Energy Physics
of the Chinese Academy of Sciences and the Institute
of Modern Physics of the Chinese Academy of Sciences and IOP Publishing Ltd}

\begin{multicols}{2}

\section{Introduction}
In tumor treatment, the carbon ion therapy have the ability to overcome the limitation of convertional radiotherapy due to its most energy deposition in selective depth, usually called Bragg peak, which results in increased biological effectiveness.
Since in carbon ion therapy, misalignment of carbon beam, patient mispositioning or changes in the structure or density of the irradiated tissue may result in dose reduction within the tumor or overdosing in organs at rish \cite{Enghardt2004284}.
So the correct depth of Bragg peak is crucial.
Considering above situation, a tool that can monitor the treatment dose distribution in vivo and non-invasive is required. 
PET scanner is a feasible technique solution for this purpose because it can image 3D distribution of positron emitters produced by nuclear fragmentation reactions of the projectiles with target nuclei~\cite{0031-9155-51-23-011,0031-9155-56-6-005}.

There three types of PET scanners~\cite{0031-9155-56-5-004} (in-beam, in-room, and off-line) have been comfirmed the feasibility to monitor the dose deposition in radioactive therapy.
Although the in-beam PET measurement is only slightly influenced by metabolic processes and blood flow, additional efforts are required for providing radiation hardness components and suppressing strong $\gamma$-ray background from the interactions of the beam and patient.
The in-room PET technique is adapted as a compatible solution between the performance and cost. In order to avoid interference between PET detectors and the hadron beam line and patient positioning, a dual-plate geometry is chosen. 

In this study, we use the GATE~\cite{0031-9155-56-4-001} simulation platform to evaluate the performances of dual-plate PET, and compare with that of four-head and full-ring PET scanners.
The dose distribution of the carbon beam and production yields of positron emitters were simulated for different carbon beam energies in a cuboid PMMA phantom.
3D images of the activity distribution of the three type scanners are produced by an iterative reconstruction alogrithm,
and the longitudinal profile images of positron emitters are compared.

\end{multicols}

\begin{figure}
    \centering
    \subfloat[Dual-plate]
    {
        \includegraphics[width=5cm]{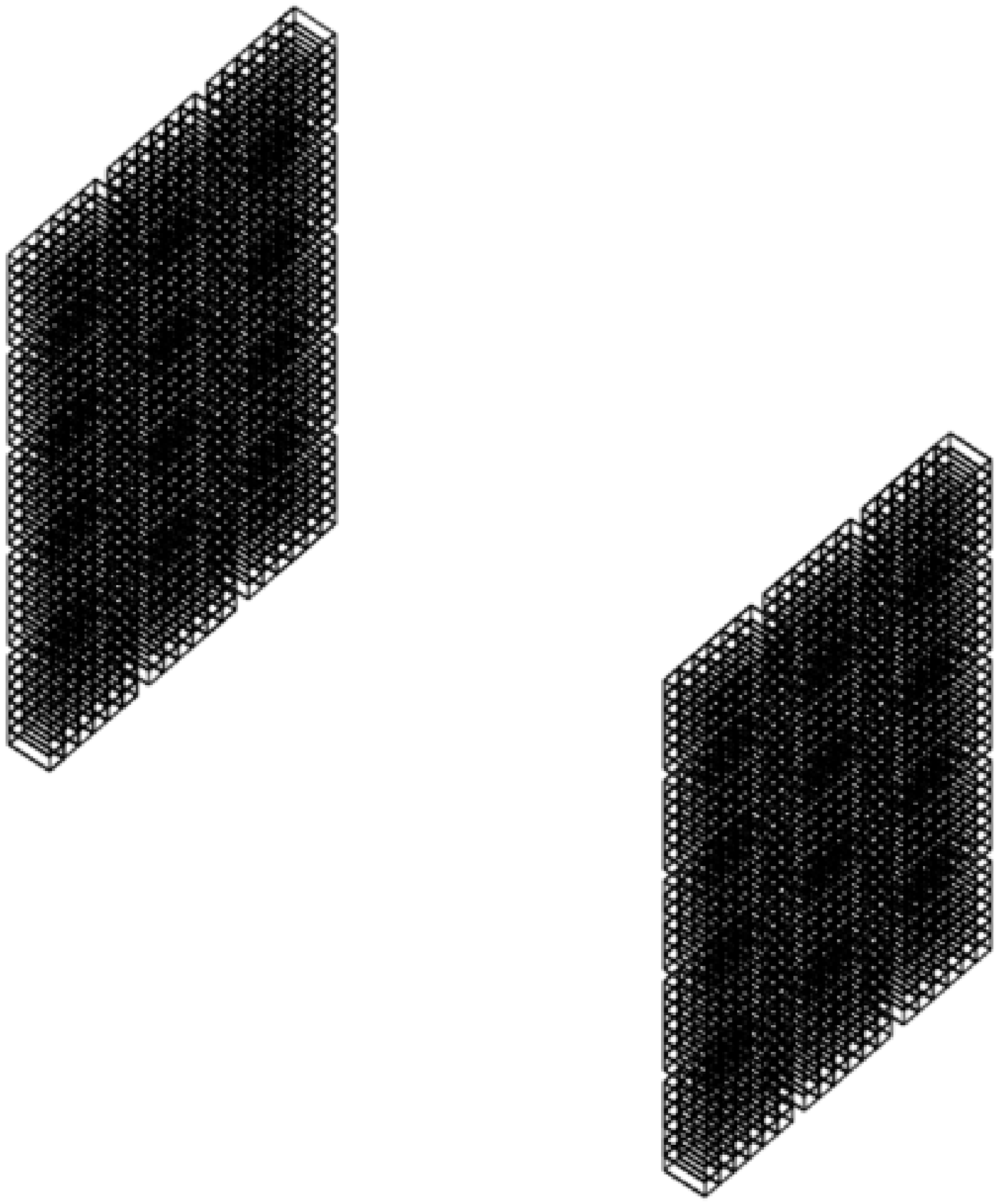}
    }
    \subfloat[Four-head]
    {
        \includegraphics[width=5cm]{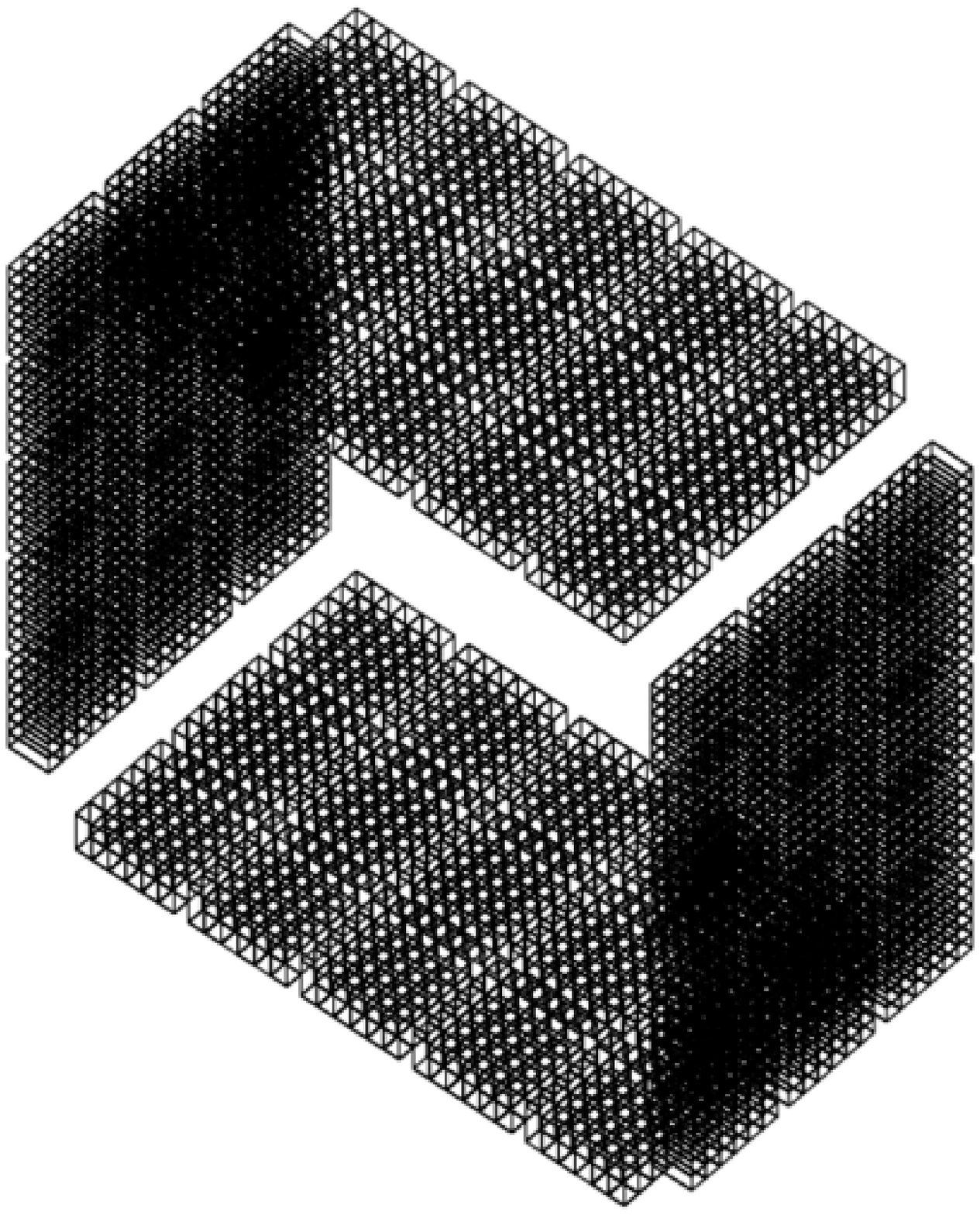}
    }
    \subfloat[Full-ring]
    {
        \includegraphics[width=5cm]{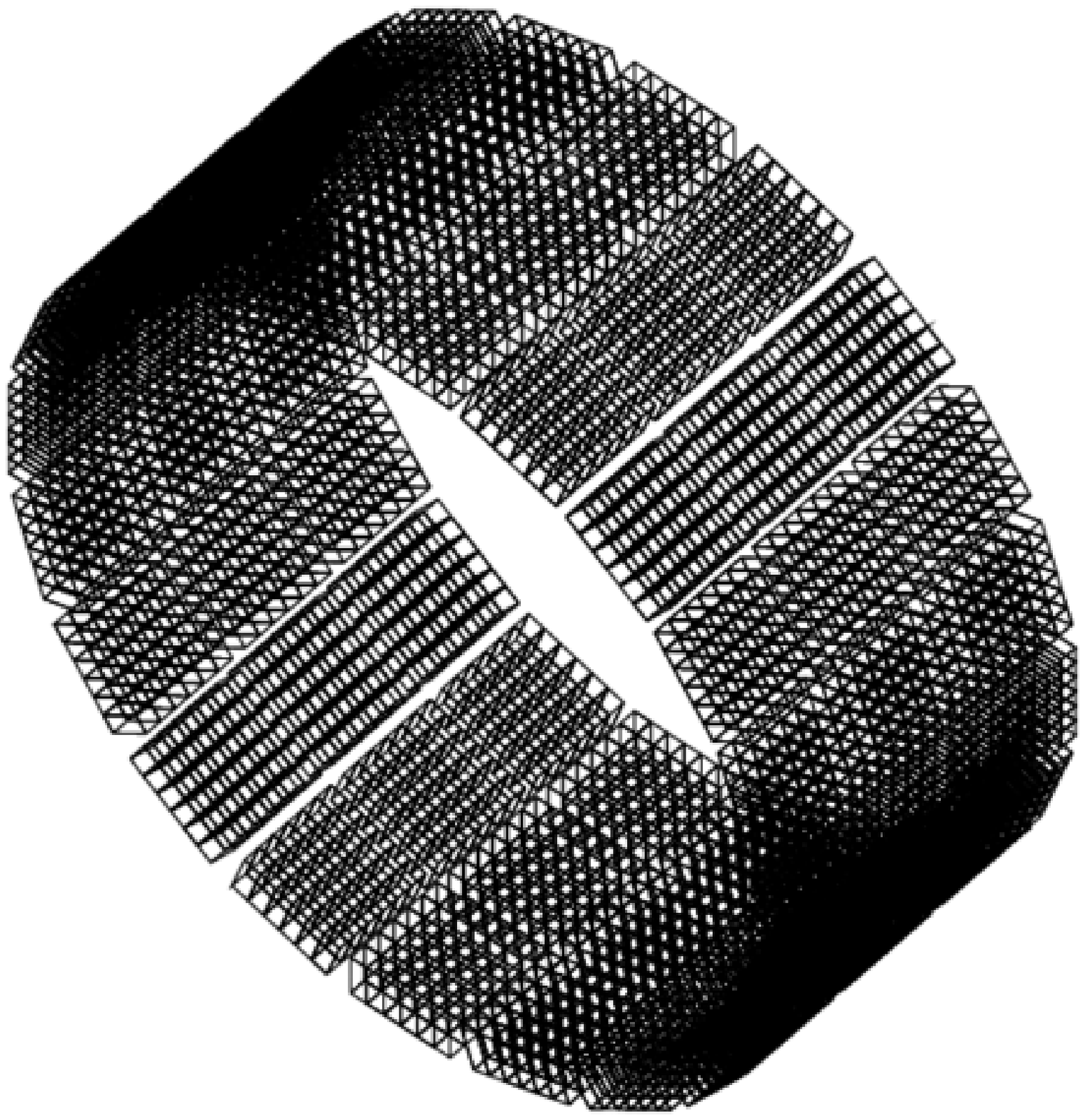}
    }
    \caption{Simplified scheme of simulated PET geometry configuration.}
    \label{fig_geometry}
\end{figure}

\begin{multicols}{2}

\section{Materials and methods}

\subsection{Description of in-room PET scanner}
To avoid the interference with beam line, the in-room PET is based on two plate heads , which are made of 3$\times$5 detector modules, respectively.
Each detector module is consisted of a 24$\times$24 array of 2$\times$2$\times$18 mm$^{3}$ LYSO pixels coupled to a Hamamatsu H8500 PMT.
Fig. \ref{fig_geometry} shows the configurations of all three simulated PET geometries.
The dual-plate, four-head and full-ring PET scanners consist of 30, 60 and 60 detector modules, respectively, with a 36 cm distance between directly opposite detector modules for dose deposition measurements.

\subsection{Performance of in-room PET scanneers}
Before radiation therapy simulation, we evaluated the spatial resolution of the reconstructed images of three kind PET scanners.
The measurement is carried out by positioning 9 $^{22}$Na point sources (0.5 mm diameter) along the X-axis,
ranging from -8 to 8 cm with 2 cm interval between each source.
The spatial resolution in the radial and tangential directions
is measured by fitting Gaussian functions to the respective profiles of the reconstructed images of the point sources.
We use the iterative reconstruction algorithm, which is based on  maximum likelihood expectation maximization (MLEM) algorithm, to produce the 3D image of the activity distribution.

\subsection{Production yield of positron emittters}

GATE V6.2, which provides lots of useful tools to collect information during simulation, is used in this study.
We use the "ProductionAndStopingActor" to estimate the distributions of $^{11}$C, $^{15}$O, $^{10}$C, and "DoseActor" to calculate the dose distributions. 
In the simulation, the carbon ion beam irradiated on a PMMA phantom with dimensions of 10$\times$15$\times$15 cm$^{3}$ for estimating the production yield and distribution of positron emitter. 
The carbon beam is delivered along Z-axis to the smallest cross-section ($10\times15~cm^{2}$) of the phantom.
Three beam energies are selected: 172, 200, 250 AMeV, according to treatment plan. The beam profile in the transverse direction is assumed to be a Gaussian shape with a FWHM of 8 mm. The intensity of the beam is 1$\times$10$^{8}$ pps. 
Positron emitters such as $^{11}$C, $^{15}$O, $^{10}$C generated in irradiated tissues by nuclear reactions are analyzed.

The phantom images are reconstructed by the MLEM algorithm, and the logitudinal profiles of the reconstructed images are calculated by ROOT software.
Last, the dose verification is evaluated by comparing the distribution of the positron emitter from "ProductionAndStoppingActor" and image profile measured by PET scanner.

\section{Results}

\subsection{Performance of 3 kinds PET scanner}

\begin{figure}
    \centering

    \subfloat[Dual-plate]
    {
        \includegraphics[width=5cm]{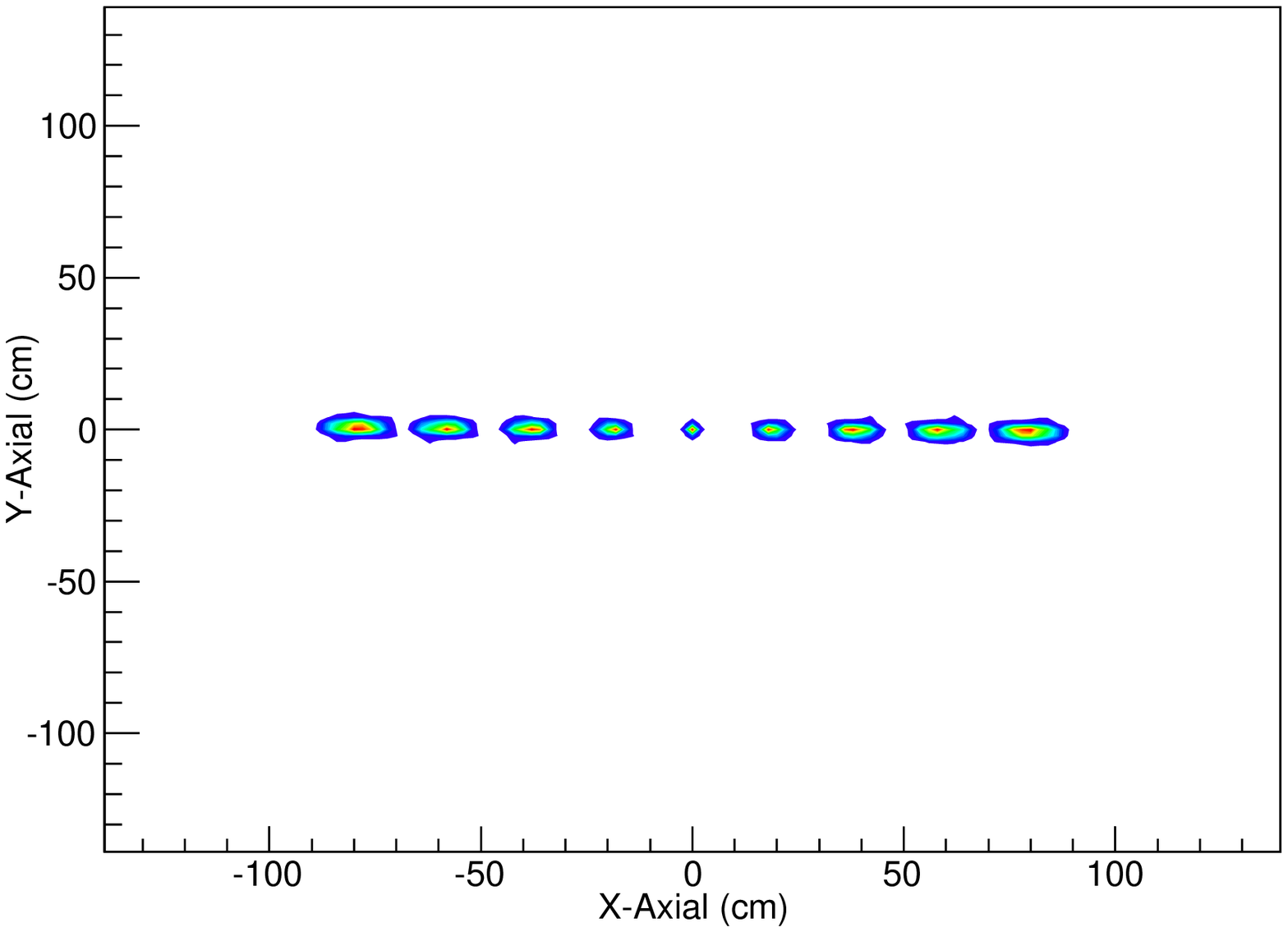}
    }
    \subfloat[Four-head]
    {
        \includegraphics[width=5cm]{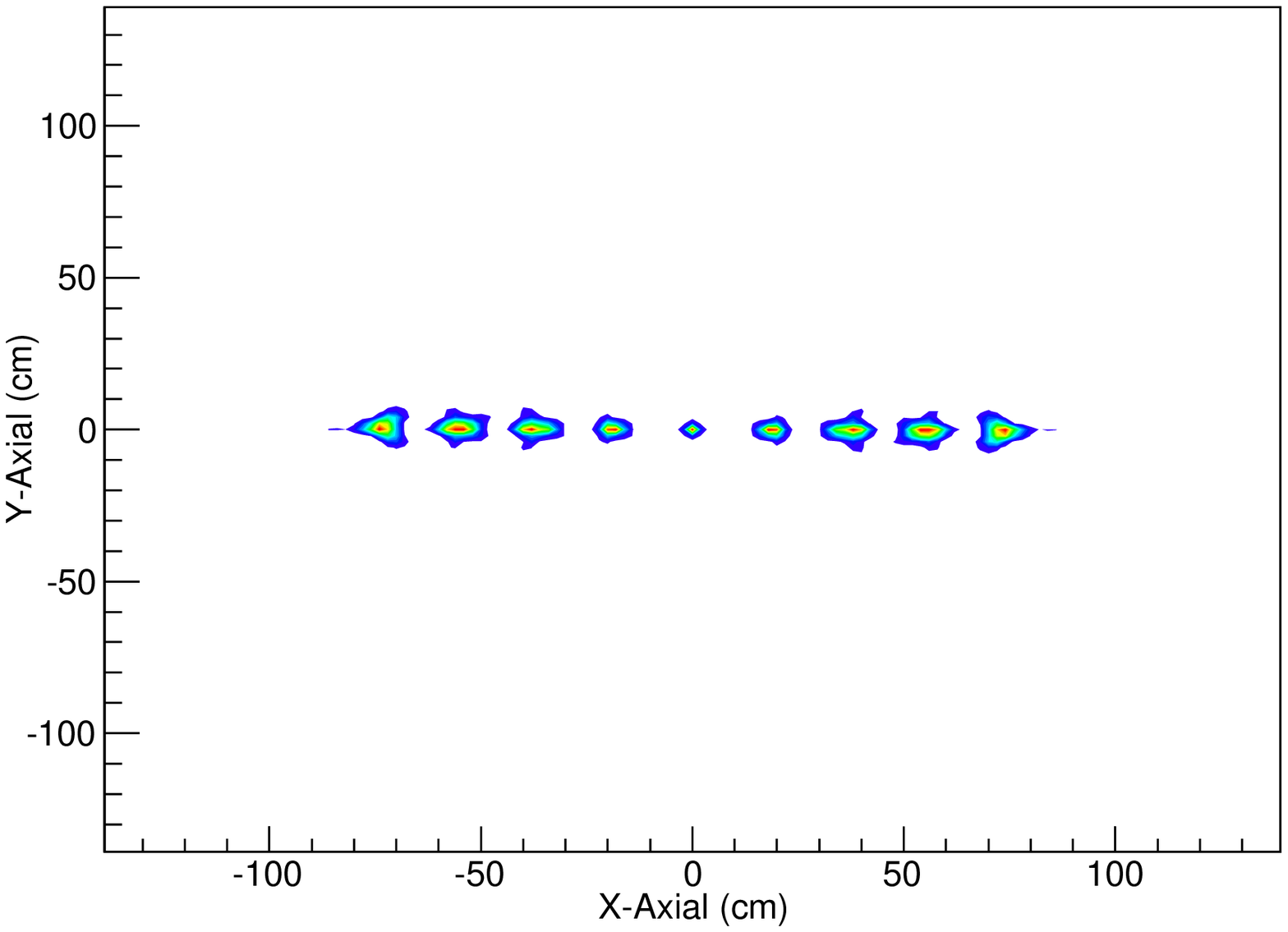}
    }
    \subfloat[Full-ring]
    {
        \includegraphics[width=5cm]{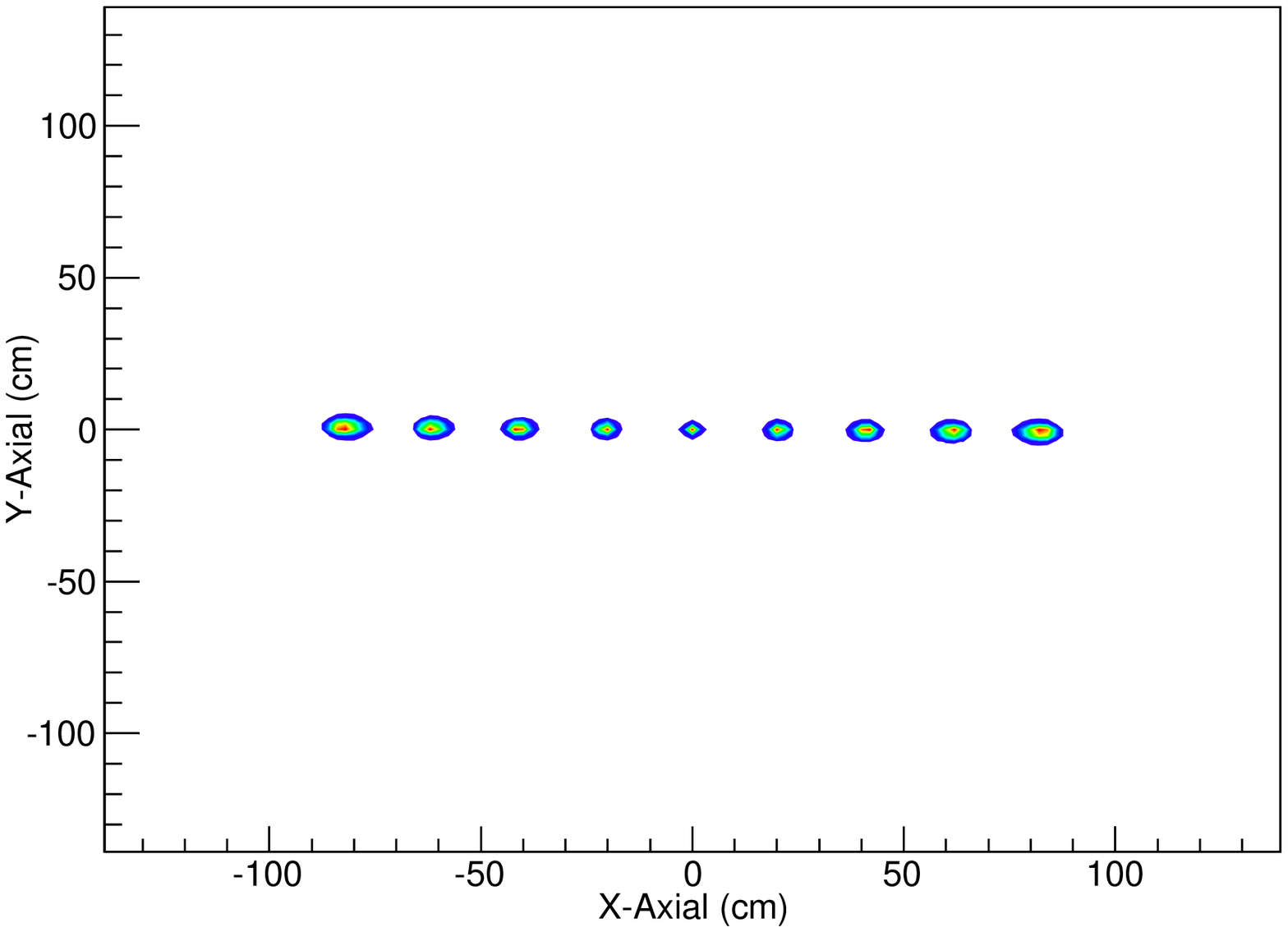}
    }

    \caption{Reconstructed images of point sources at different positions}
    \label{fig_point_source}
\end{figure}

Fig.\ref{fig_point_source} shows reconstructed images of points sources at different positions measured by the three PET scanners.
With the dual-head, image of the point sources in the near-peripheral region of FOV is blurred, while the full-ring PET scanner shows relative uniform imaging characteristics over the entir FOV.
The radial and tangential spatial resolutions are illustrated in Fig.\ref{fig_resolution} for different position along the X-axis across the FOV.

\end{multicols}

\begin{multicols}{2}

\begin{center}
    \includegraphics[width=6cm]{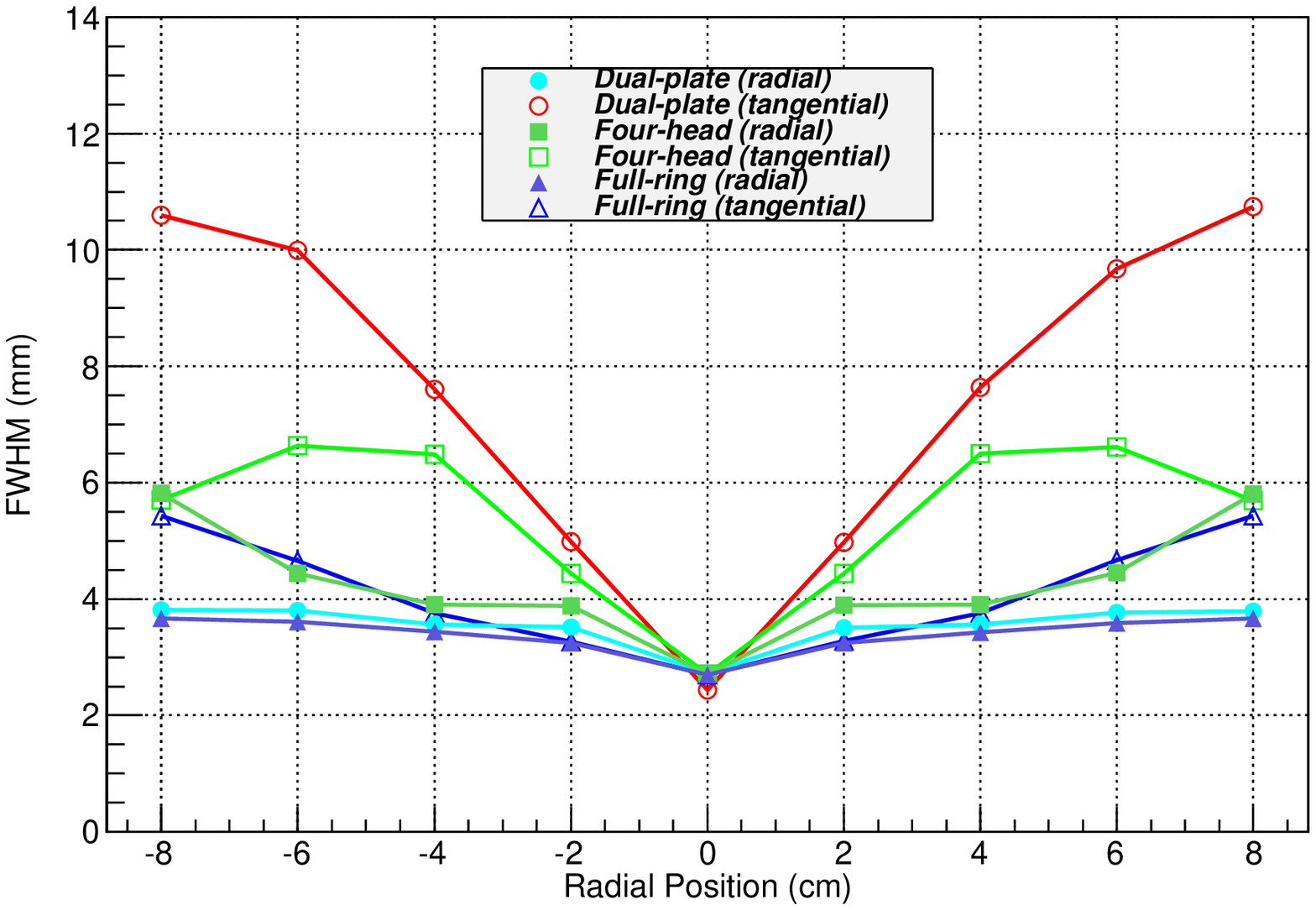}
    \figcaption{\label{fig_resolution} Spatial resolution in radial and tangential directions for radial locations in the FOV. Resolutions are based on a 1.0 MBq $^{22}$Na point source measured in air and reconstructed with MLEM.} 
\end{center}

\begin{center}
    \includegraphics[width=6cm]{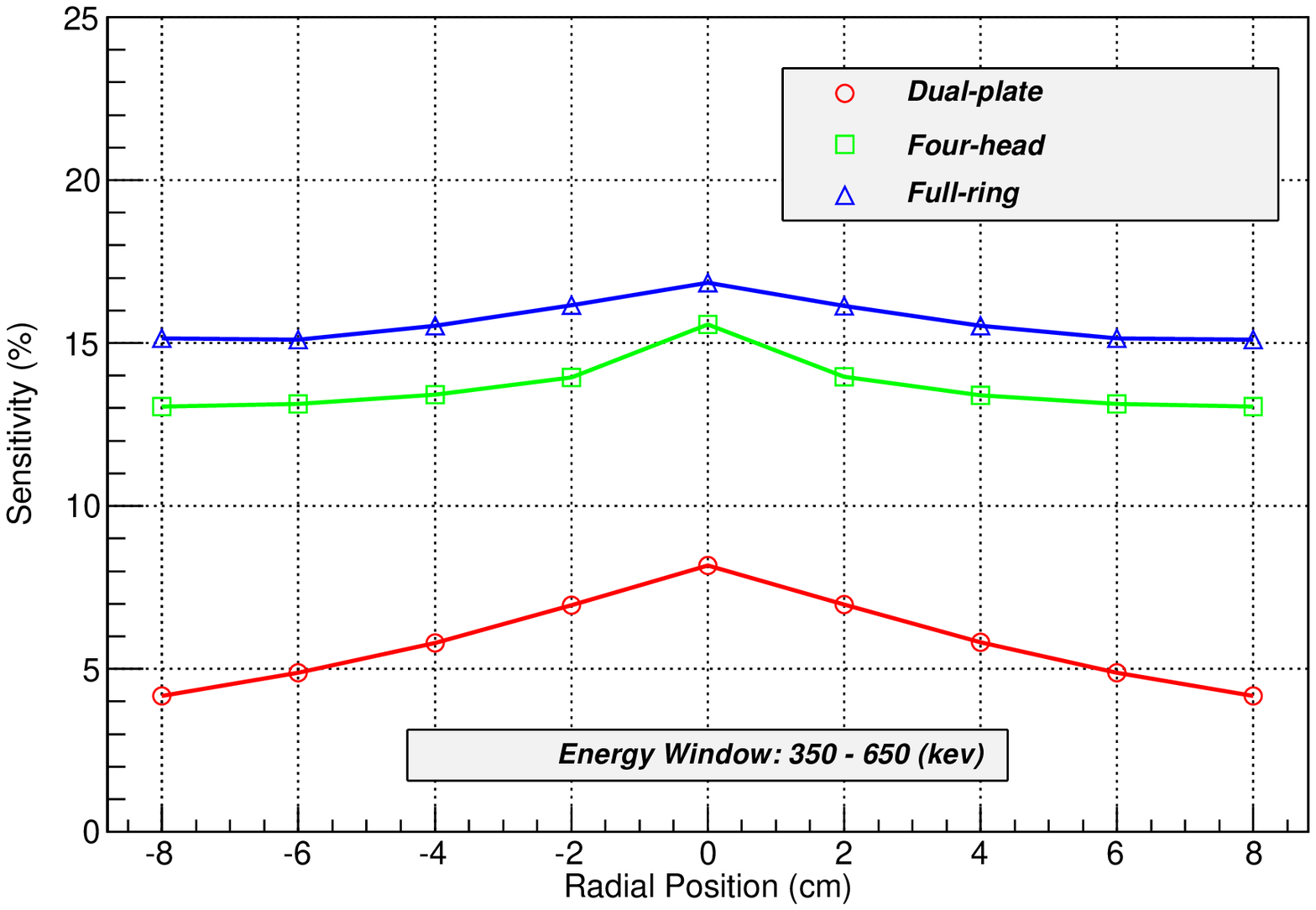}
    \figcaption{\label{fig_sensitivity} The sensitivities were measured with a 1.0 MBq $^{22}$Na point source, stepped at 2-cm increments in radial direction and with an energy window (350 - 650 keV).}
\end{center}
\end{multicols}

\begin{multicols}{2}
The sensitivities of 3 kinds of PET scanner, measured with a 1.0 MBq $^{22}$Na point source, stepped at 2-cm increments in radial direction and with an energy window (350 - 650 keV), are shown in Fig.~\ref{fig_sensitivity}. The sensitivity of ring is nearly three times higher than that of the dual-plate.

\subsection{Yields and distributions of positron emitters}
The yields of $^{11}$C, $^{15}$O and $^{10}$C produced by 172, 200, 250 AMeV carbon beams, respectively, in the PMMA cuboid phantom are simulated, and listed in Table.\ref{tabl1}.
The yield of $^{11}$C, which almost dominates the contribution of positron emitters, is 6 times higher than those of $^{15}$O and $^{10}$C.

\begin{center}
\tabcaption{ \label{tabl1} Calculated yields of positron-emitting nuclei produced by 172, 200 and 250 AMeV $^{12}$C ions.}
\footnotesize
\begin{tabular*}{80mm}{c@{\extracolsep{\fill}}rrrr}
\toprule 
&172 AMeV&200 AMeV&250 AMeV\\
\hline
$^{11}$C&6.88 \%&8.99 \%&12.03 \% \\
$^{10}$C&0.96 \%&1.10 \%&1.59 \% \\
$^{15}$O&1.21 \%&1.45 \%&2.23 \% \\
\bottomrule
\end{tabular*}
\end{center}

Fig.\ref{fig_yield} shows the spatial distribution of positron emitters with the corresponding depth-dose distribution of carbon beam.
The distance to 50\% distal falloff of Bragg peak is 56, 72, 106 mm for the 172, 200, 250 AMeV carbon beams, respectively, and the relative distances between 50\% distal falloff of Bragg peak and that of positron activity peak are 1.2\%, 1.4\% and 1.0\%.

\end{multicols}
\begin{center}
    (a)
    {
        \includegraphics[width=5cm]{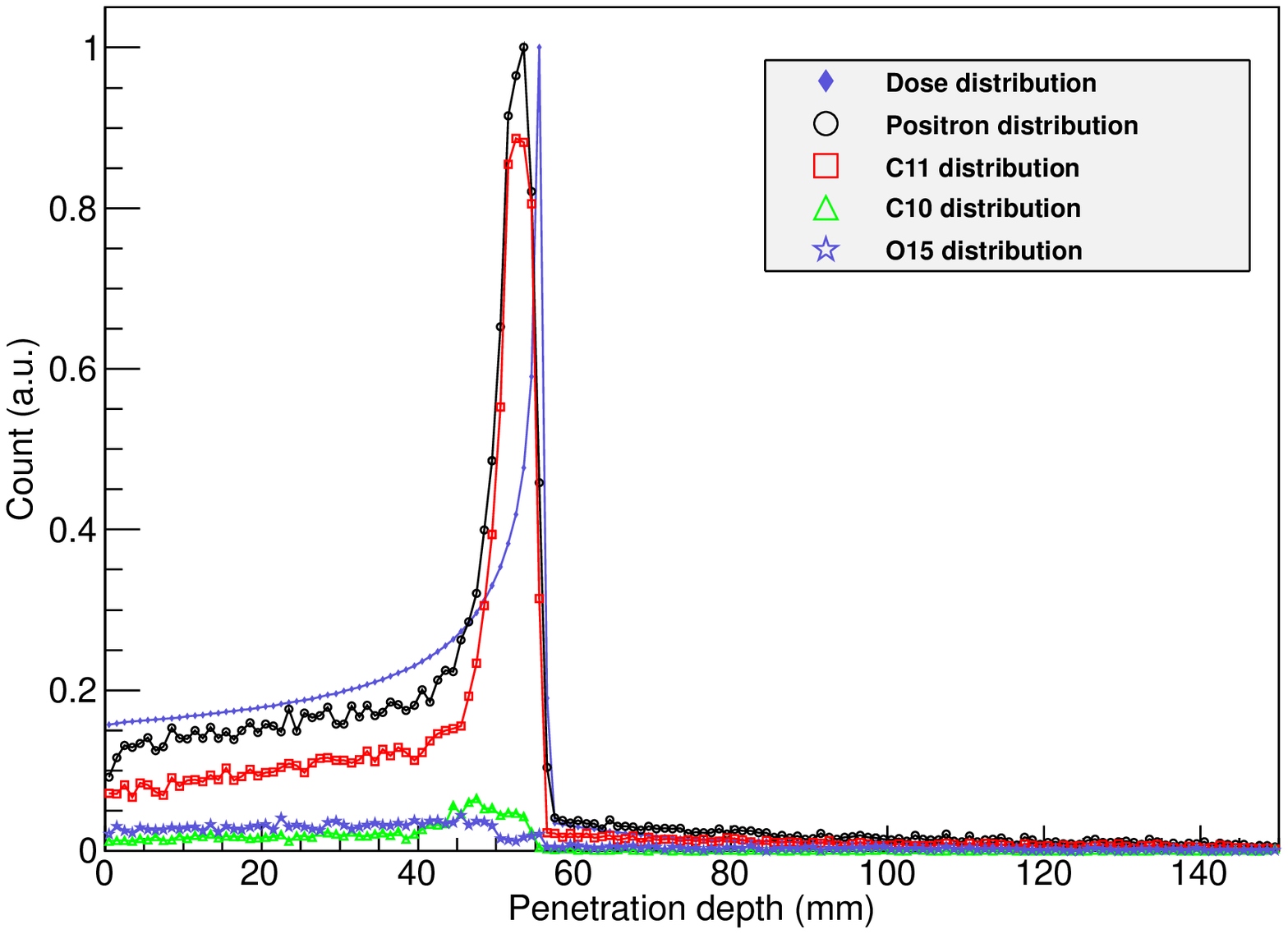}
    }
    (b)
    {
        \includegraphics[width=5cm]{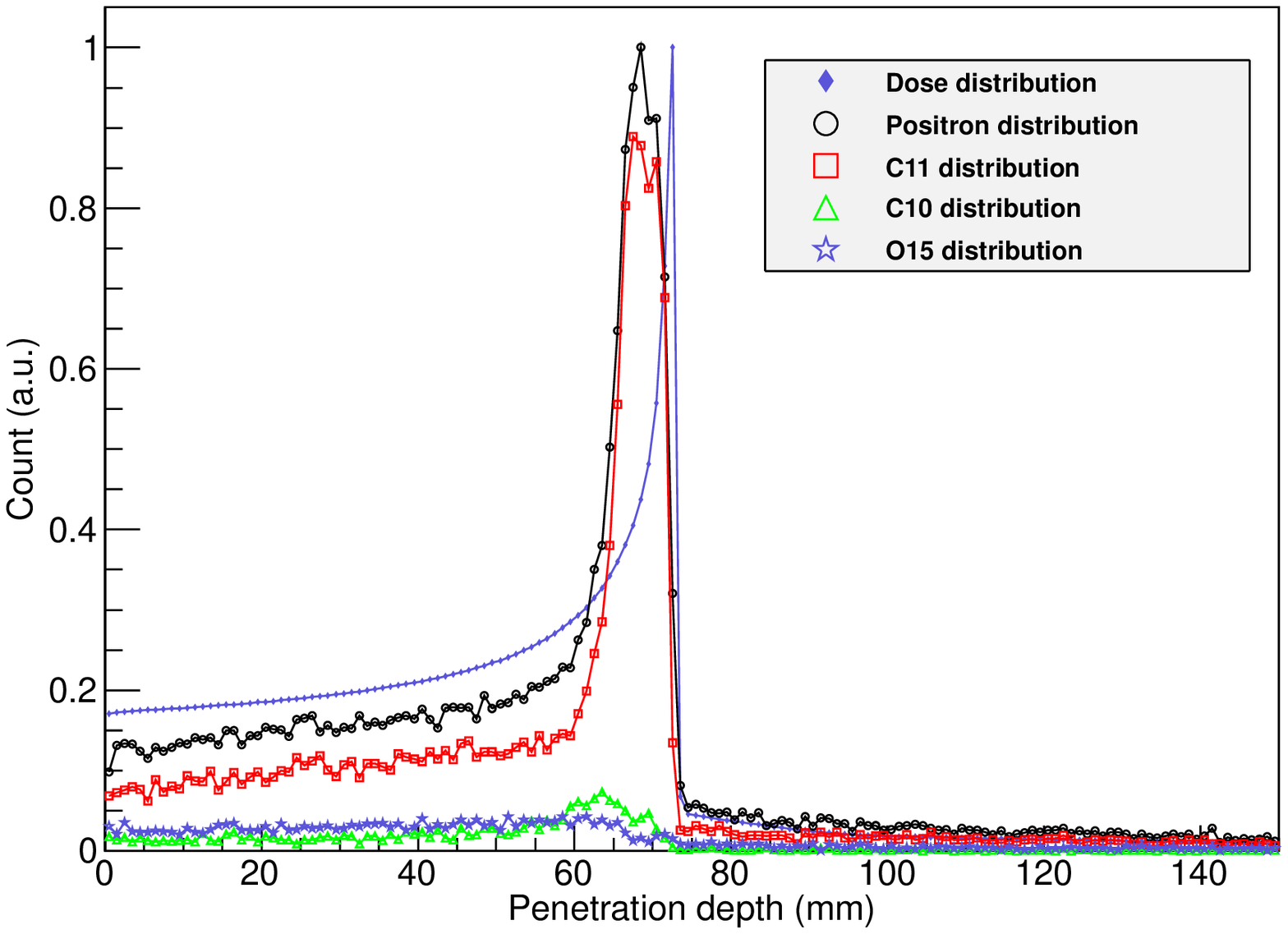}
    }
    (c)
    {
        \includegraphics[width=5cm]{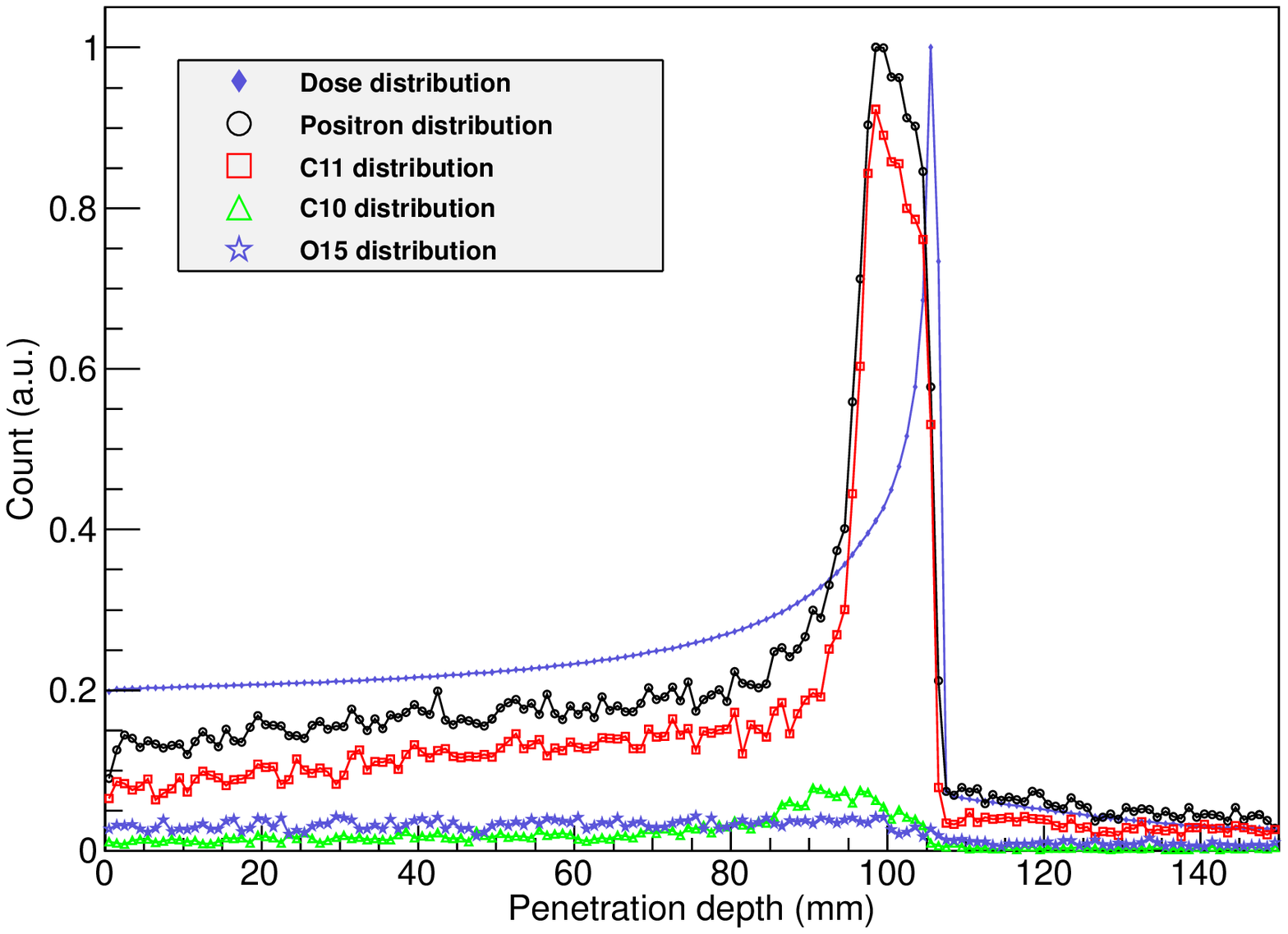}
    }
    \figcaption{\label{fig_yield} Simulated depth distribution of the deposited energy (diomand) and positron activity (circle) for (a) 172, (b) 200, (c) 250 AMeV $^{12}$C nuclei in the PMMA phantom. The distribution of $^{11}$C, $^{10}$C and $^{15}$O are shown by square, triangle and star, respectively.}
\end{center}

\begin{figure}[!ht]

\begin{center}
\captionsetup[subfloat]{labelformat=parens}    
    \sidesubfloat[]
    {
        \includegraphics[width=3.8cm]{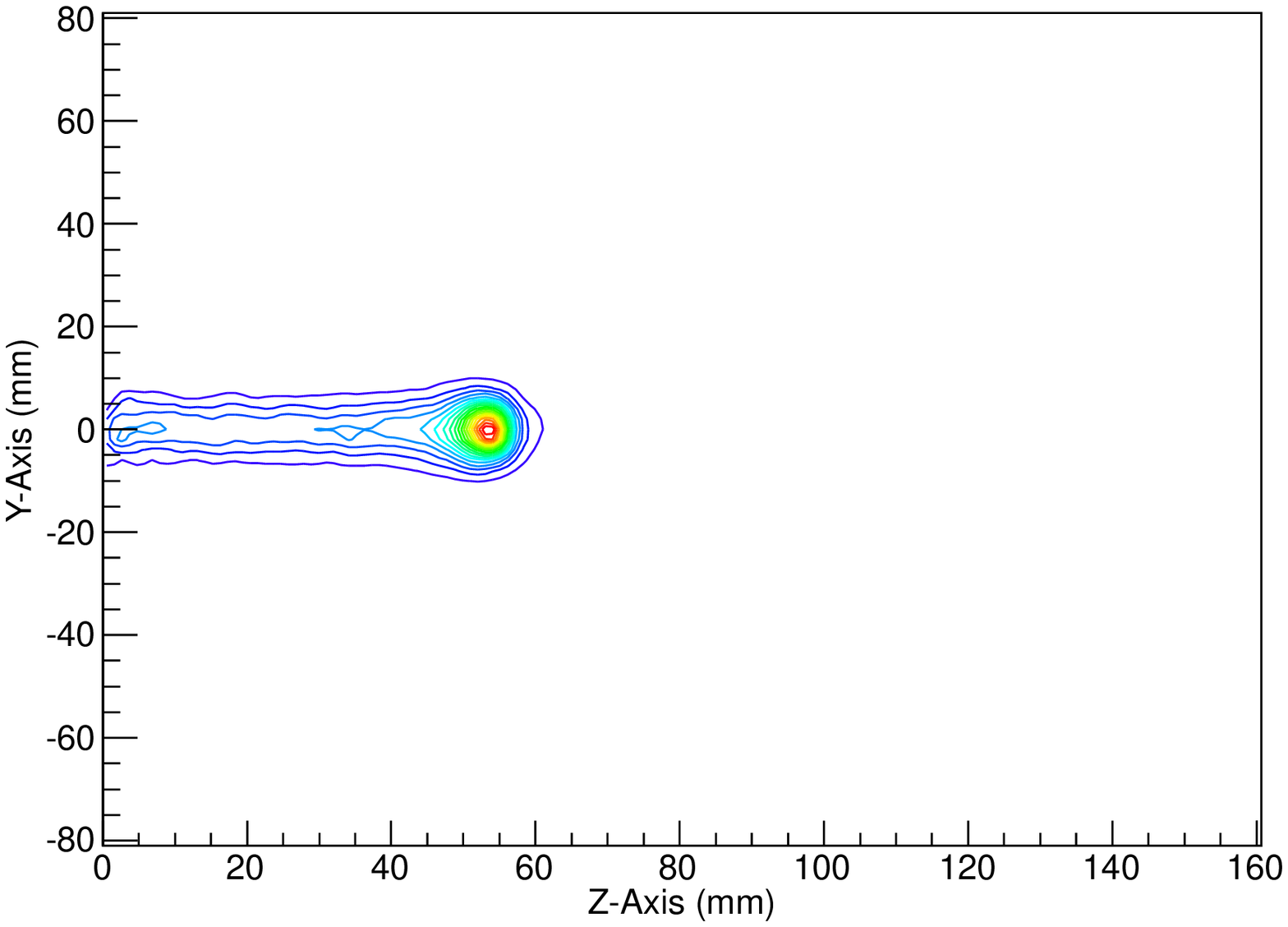}
    }
\captionsetup[subfloat]{labelformat=empty}    
    \subfloat
    {
        \includegraphics[width=3.8cm]{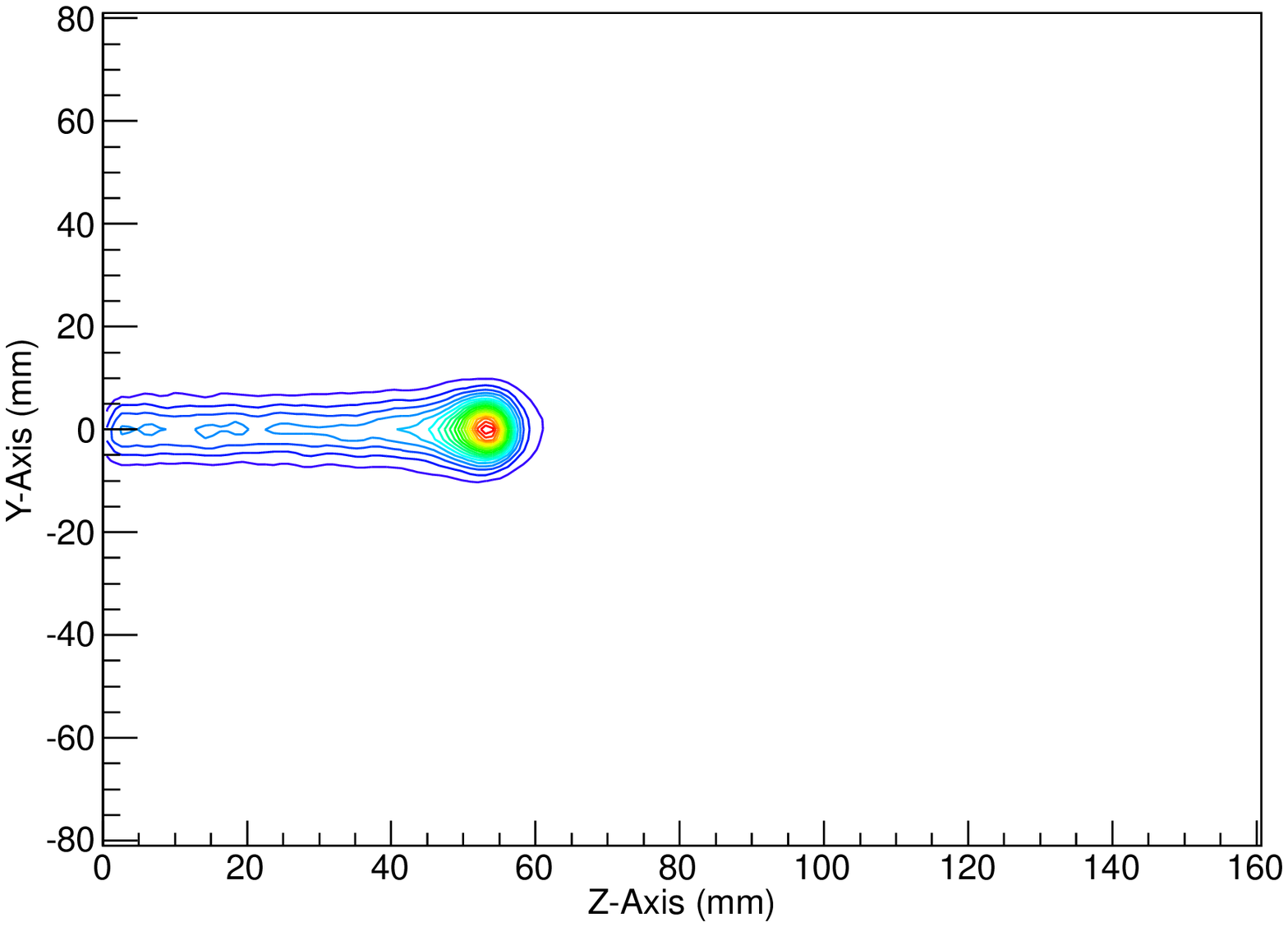}
    }
    \subfloat
    {
        \includegraphics[width=3.8cm]{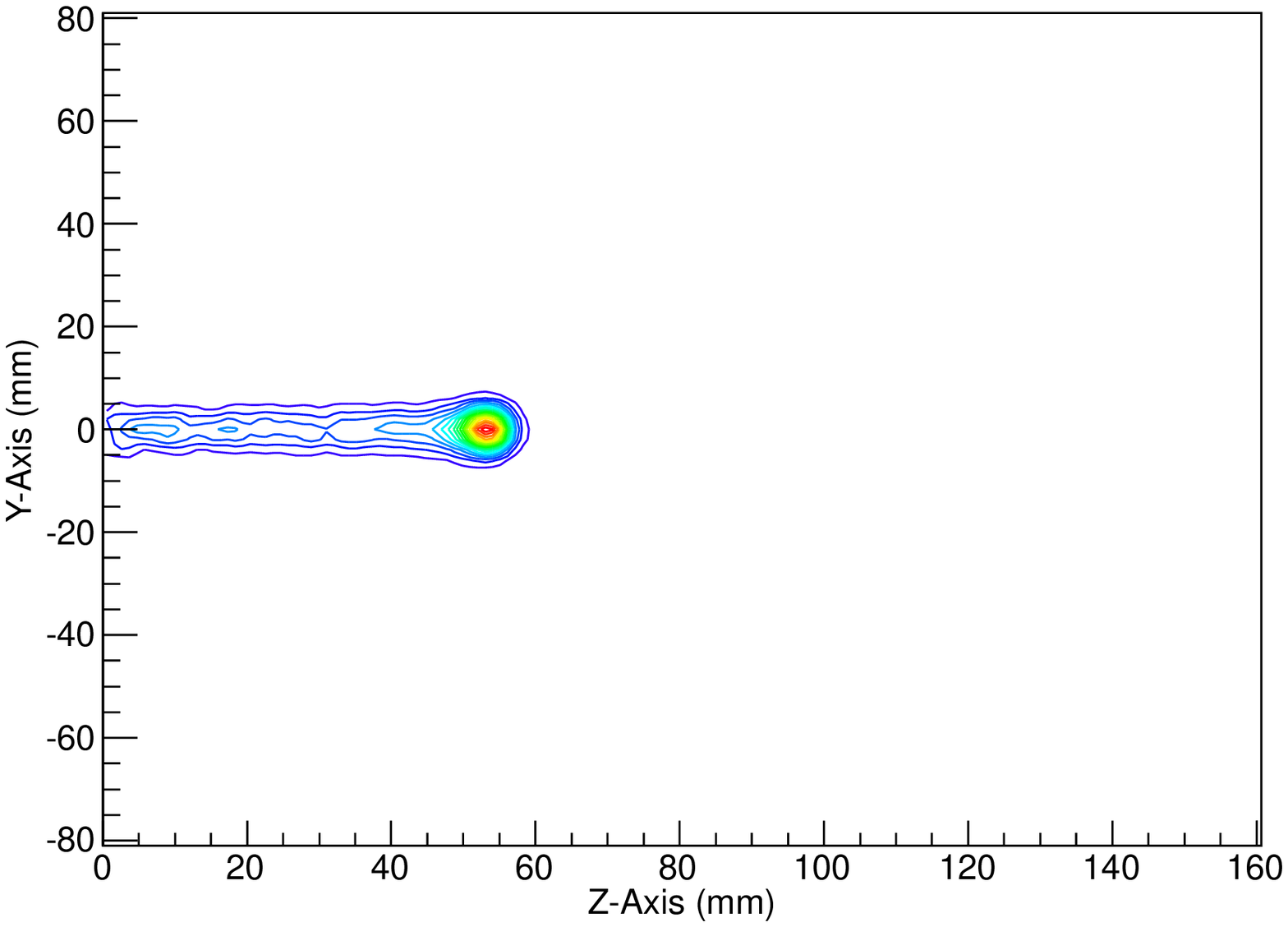}
    }
\\
    \subfloat[Dual-plate]
    {
        \includegraphics[width=3.8cm]{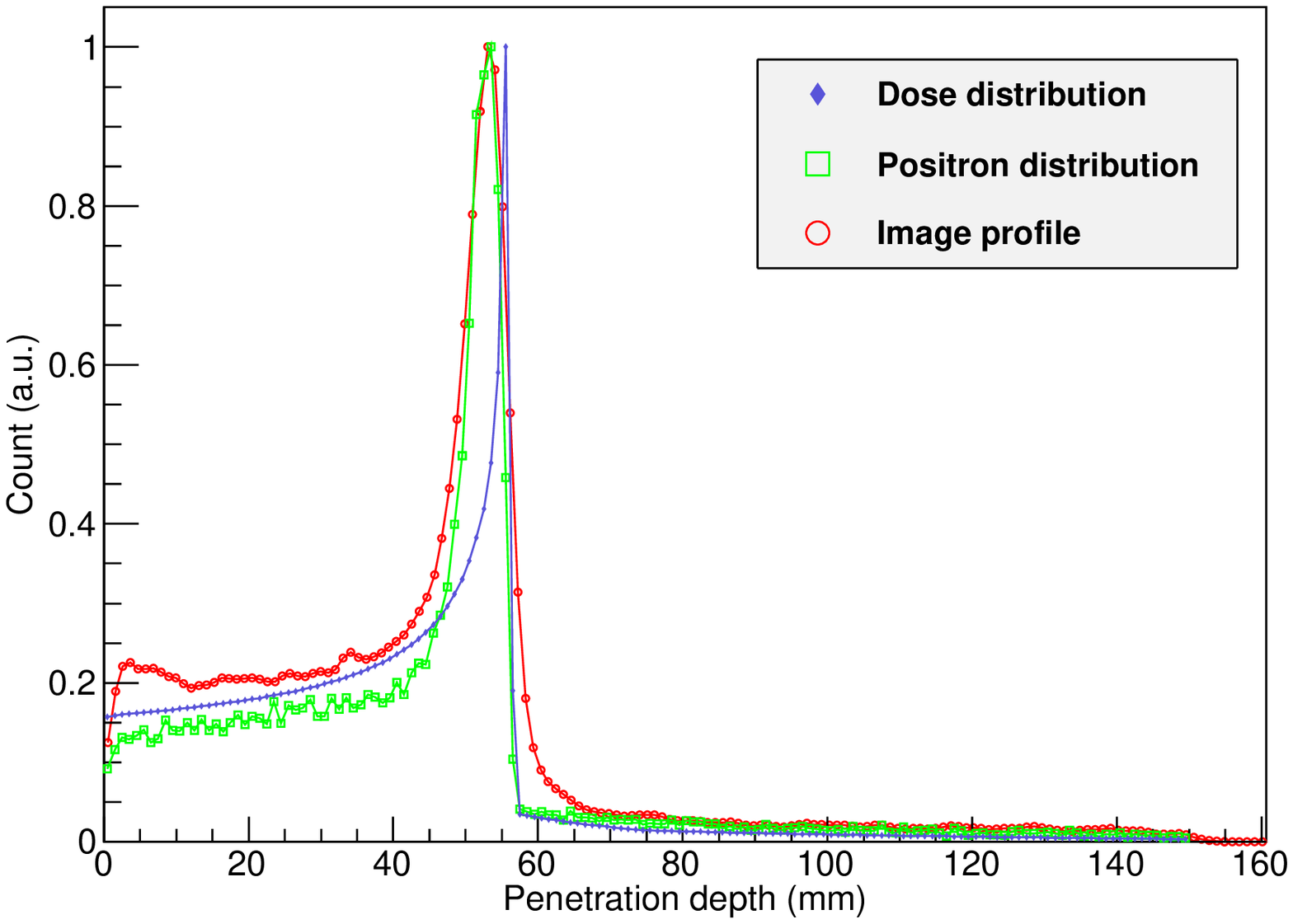}
    }
    \subfloat[Four-head]
    {
        \includegraphics[width=3.8cm]{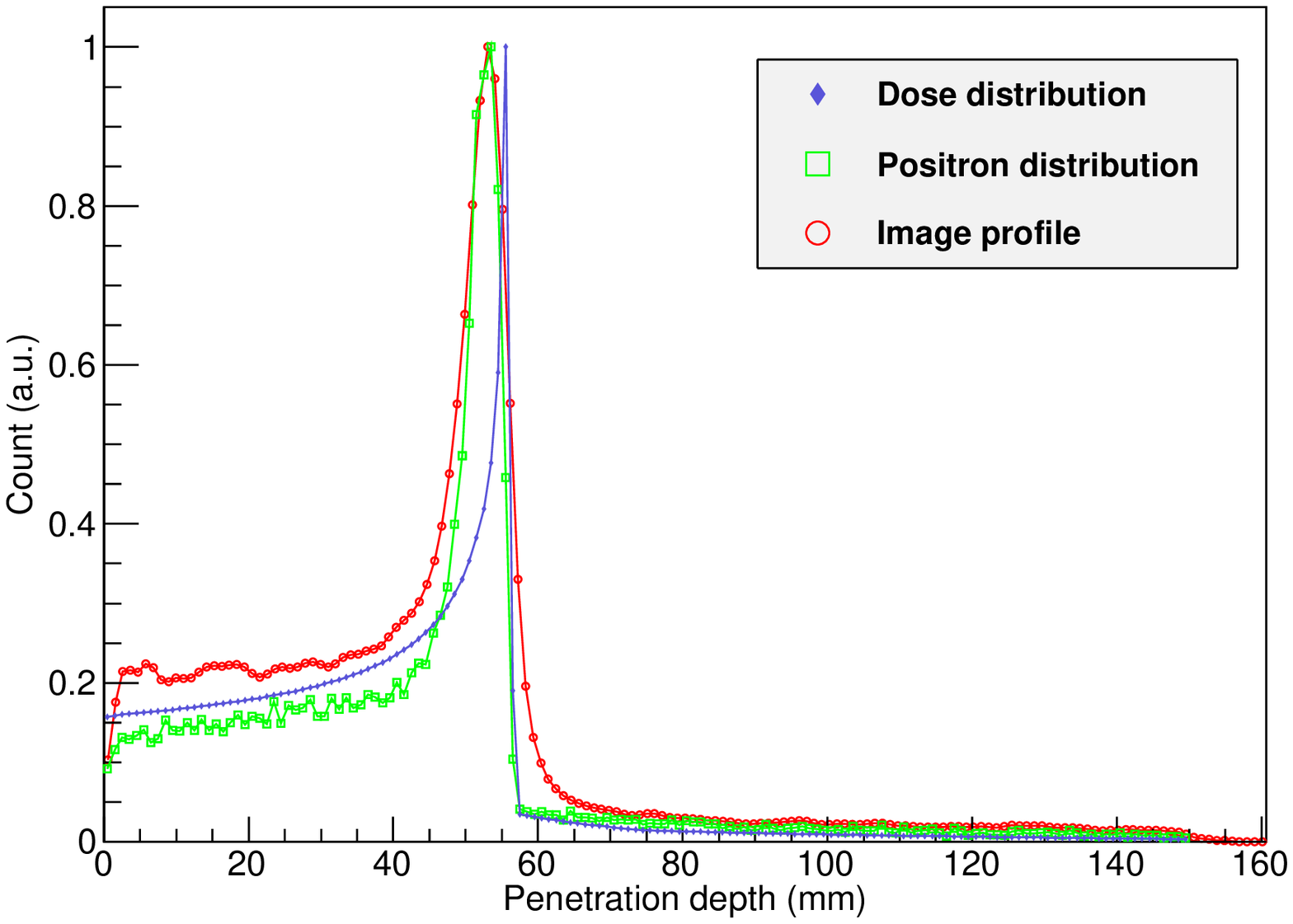}
    }
    \subfloat[Full-ring]
    {
        \includegraphics[width=3.8cm]{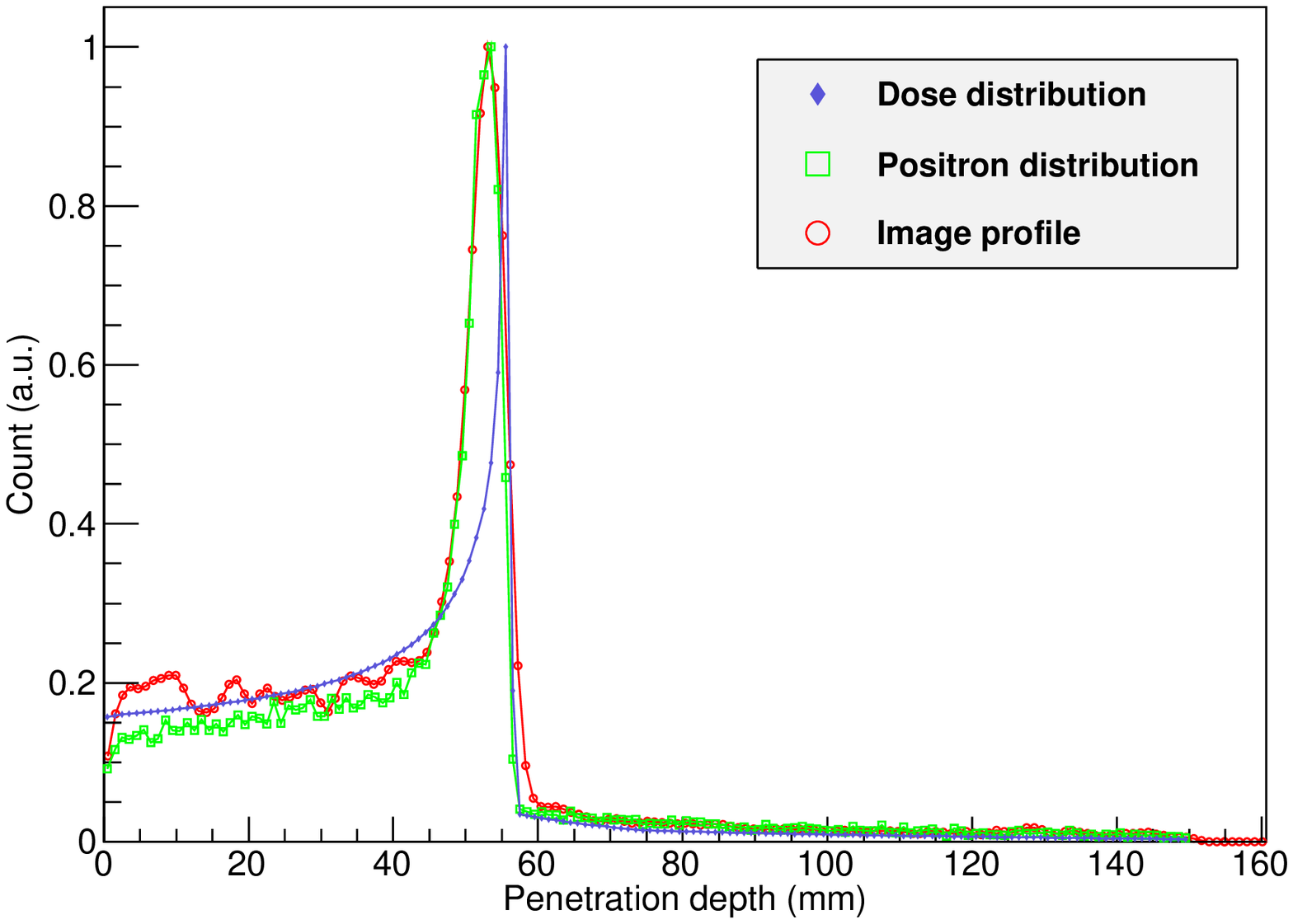}
    }
\end{center}


\begin{center}
\captionsetup[subfloat]{labelformat=parens}
\setcounter{subfigure}{1}  
   \sidesubfloat[]
    {
        \includegraphics[width=3.8cm]{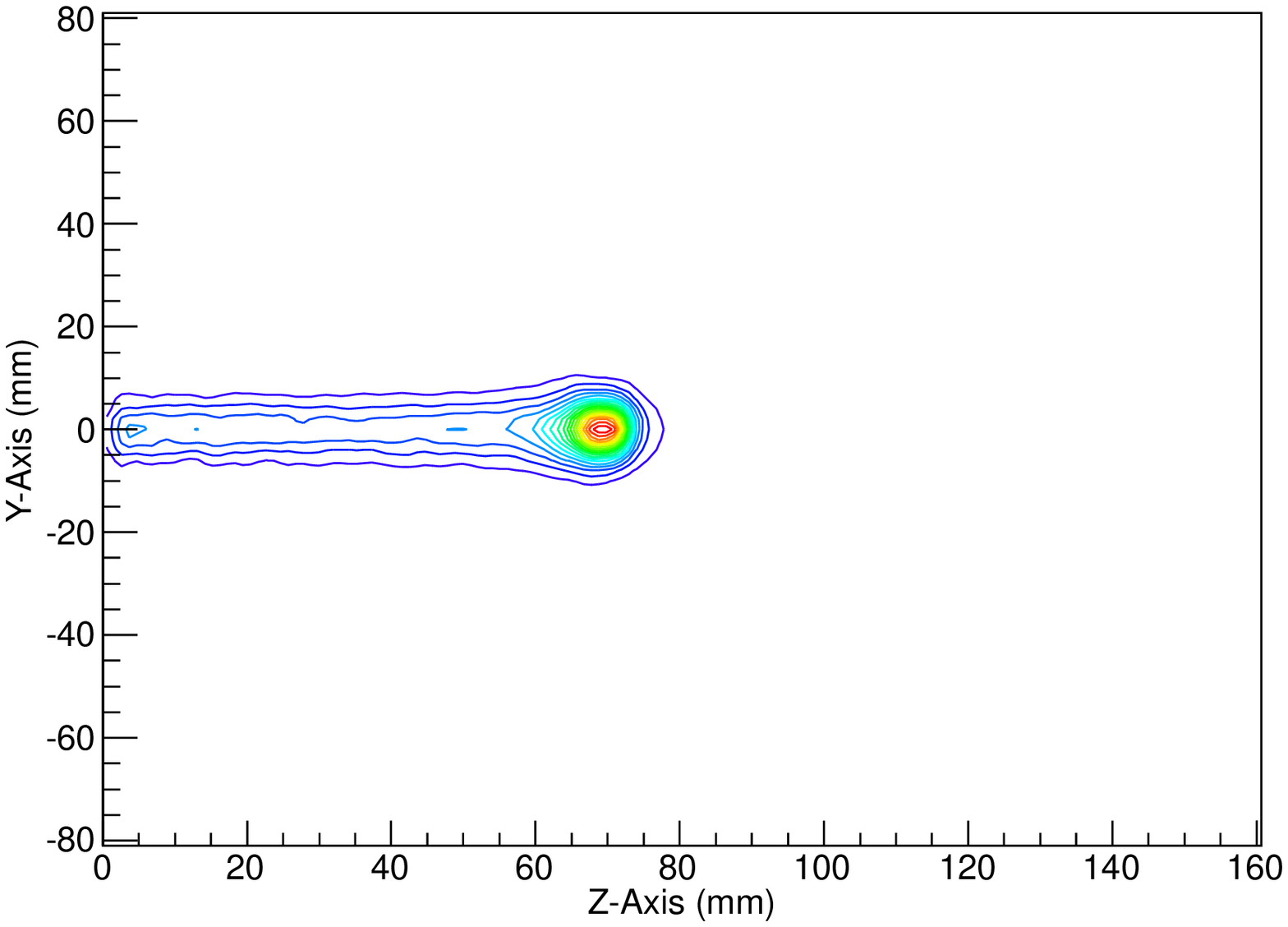}
    }
\captionsetup[subfloat]{labelformat=empty} 
    \subfloat
    {
        \includegraphics[width=3.8cm]{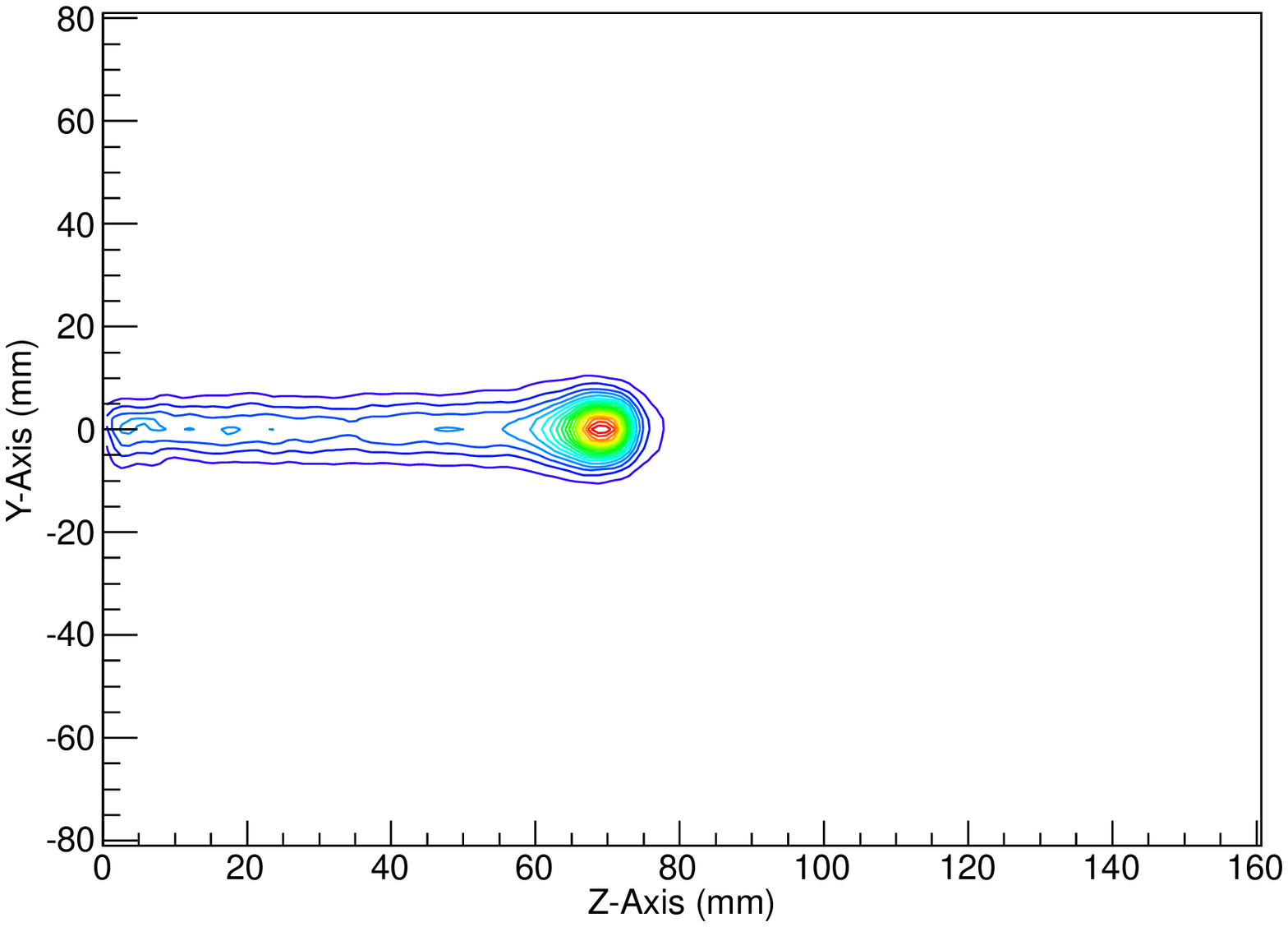}
    }
    \subfloat
    {
        \includegraphics[width=3.8cm]{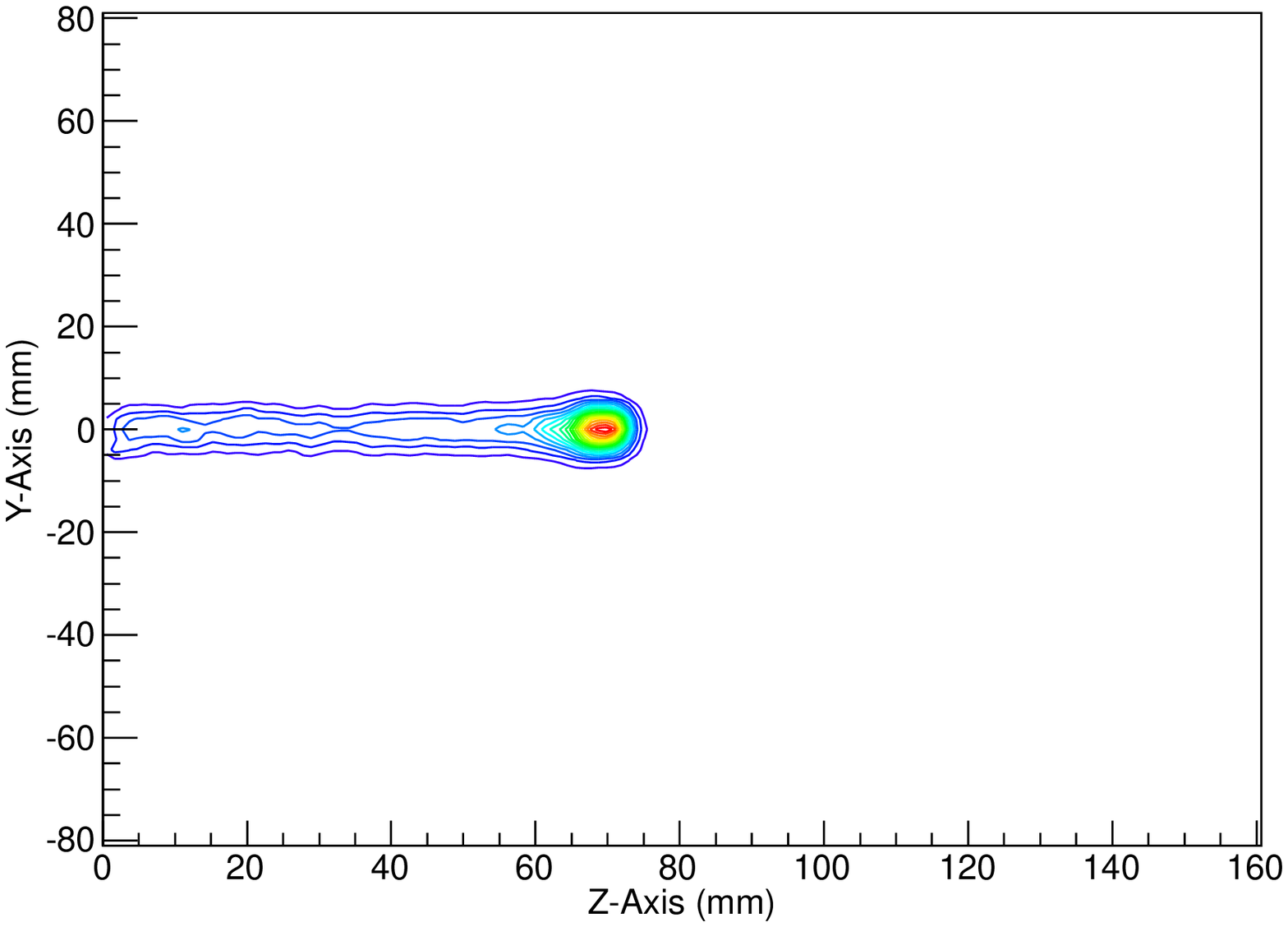}
    }
\\
    \subfloat[Dual-plate]
    {
        \includegraphics[width=3.8cm]{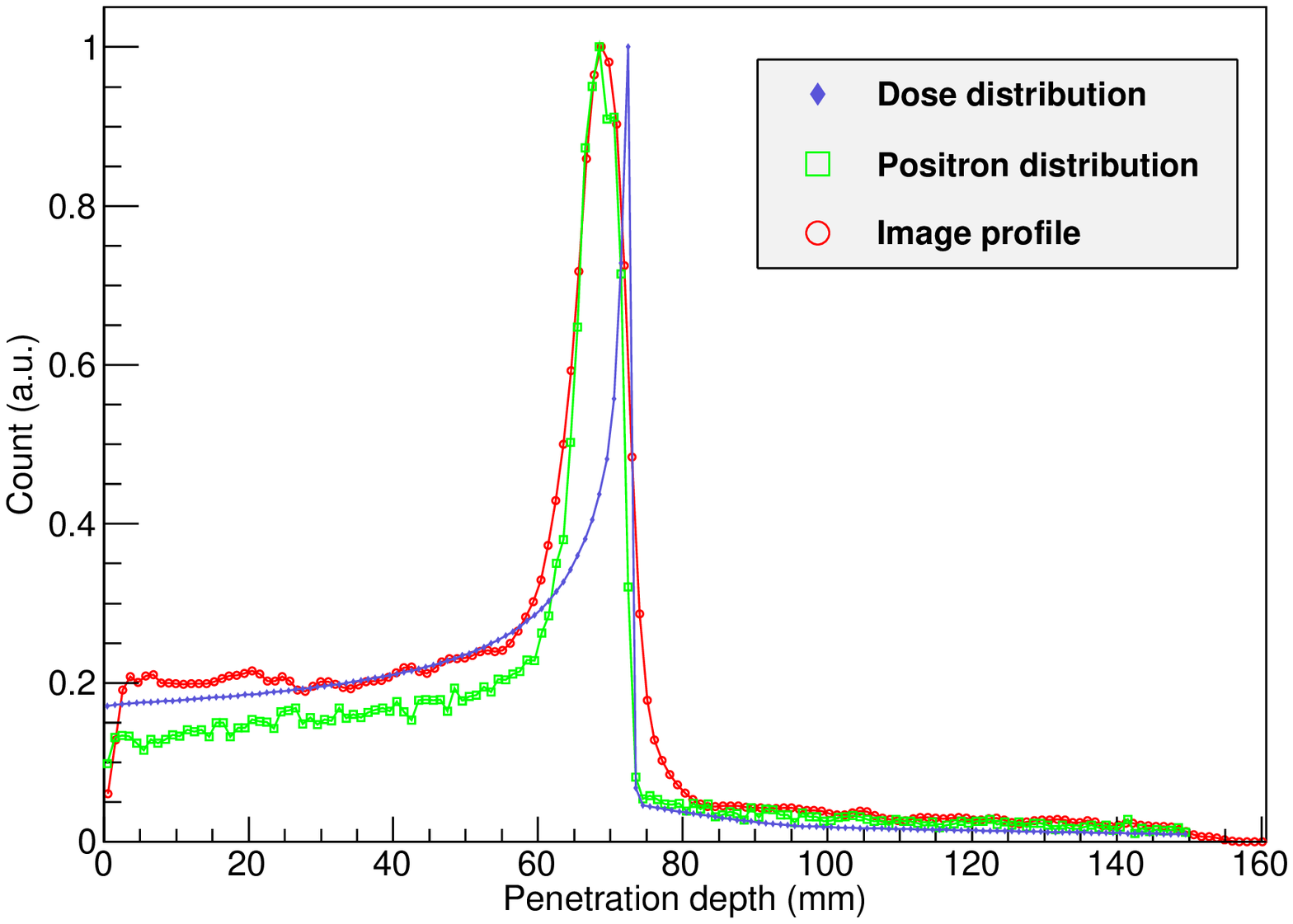}
    }
    \subfloat[Four-head]
    {
        \includegraphics[width=3.8cm]{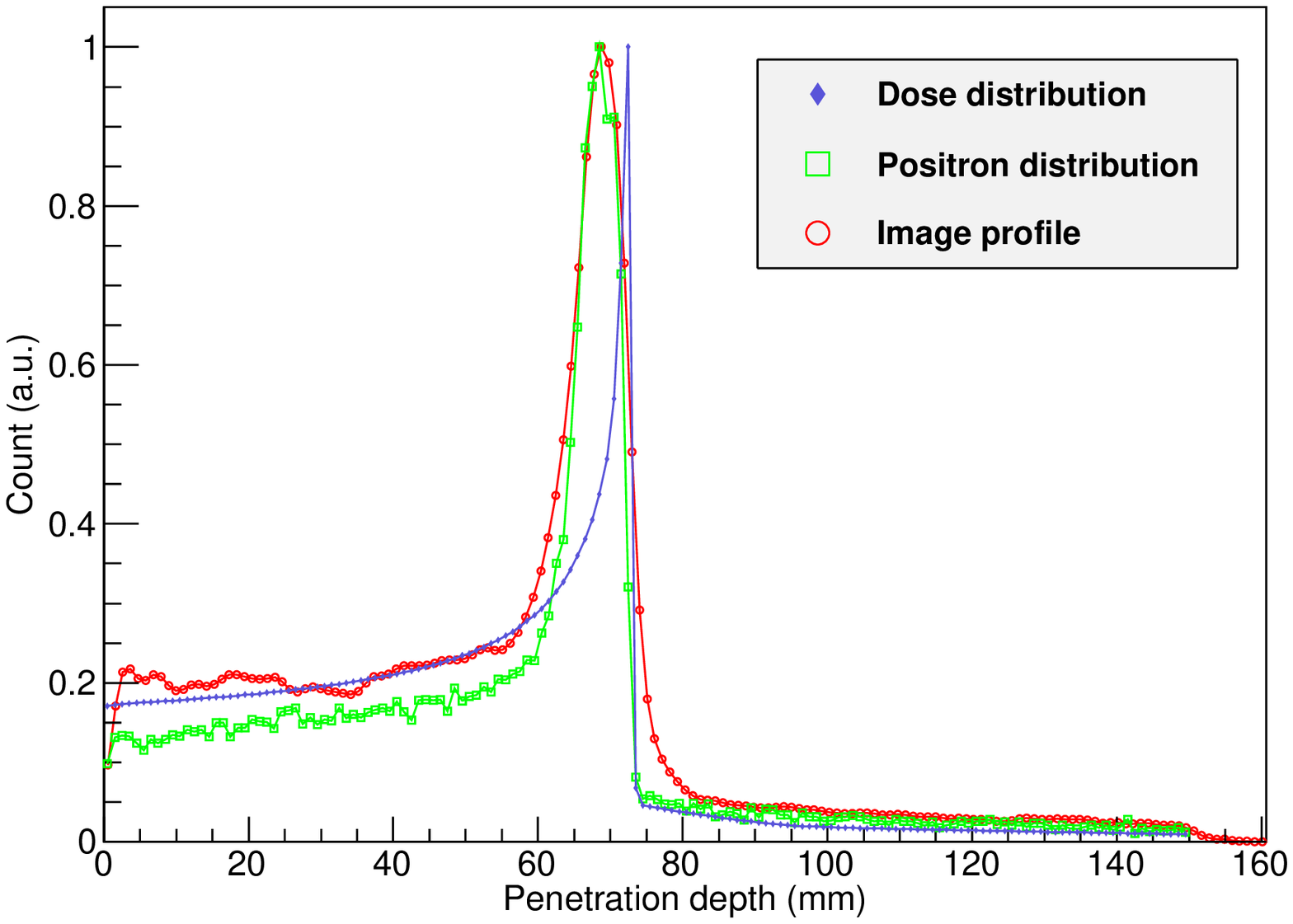}
    }
    \subfloat[Full-ring]
    {
        \includegraphics[width=3.8cm]{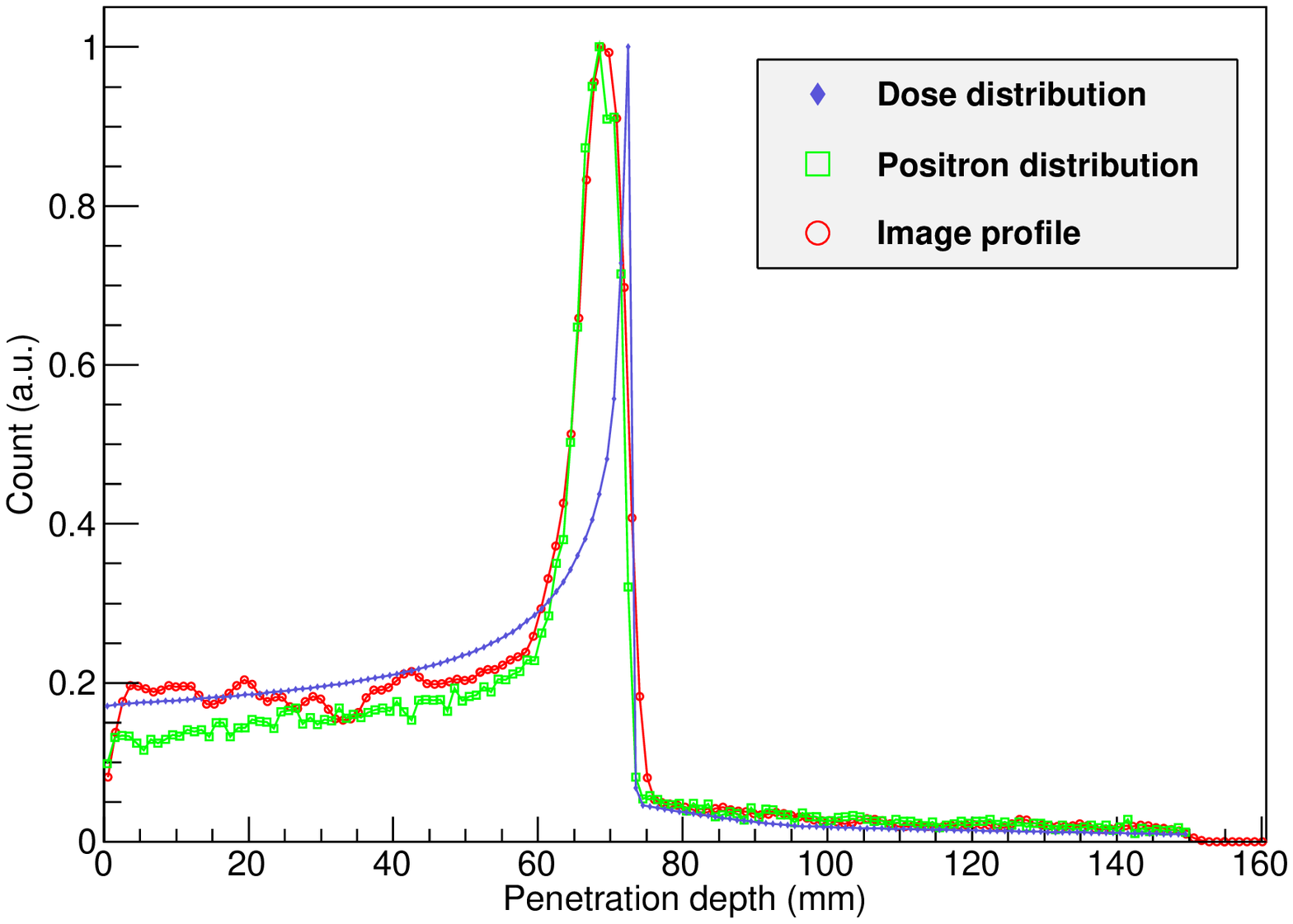}
    }
\end{center}


\begin{center}
\captionsetup[subfloat]{labelformat=parens}  
\setcounter{subfigure}{2}  
   \sidesubfloat[]
    {
        \includegraphics[width=3.8cm]{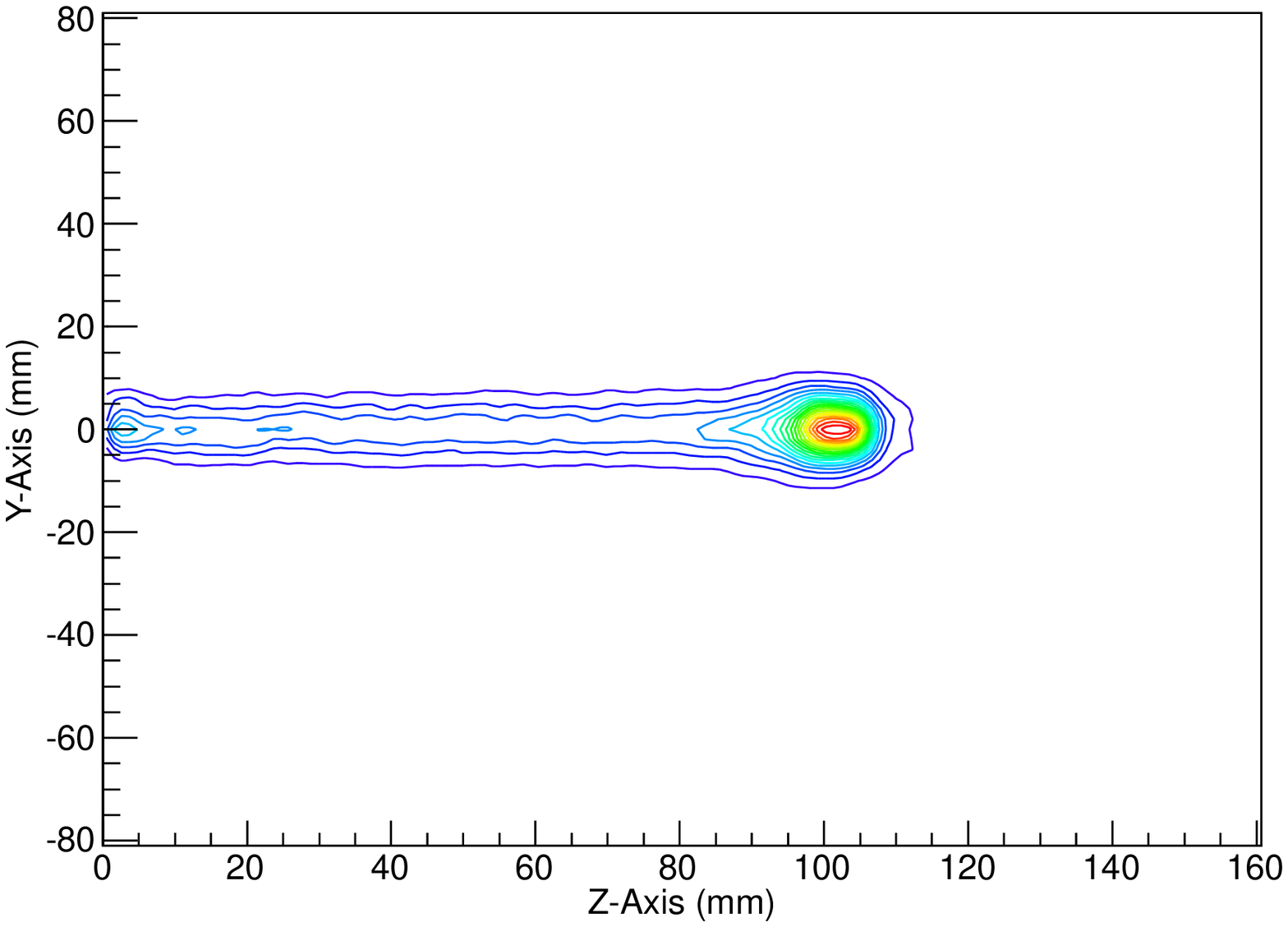}
    }
\captionsetup[subfloat]{labelformat=empty}     
    \subfloat
    {
        \includegraphics[width=3.8cm]{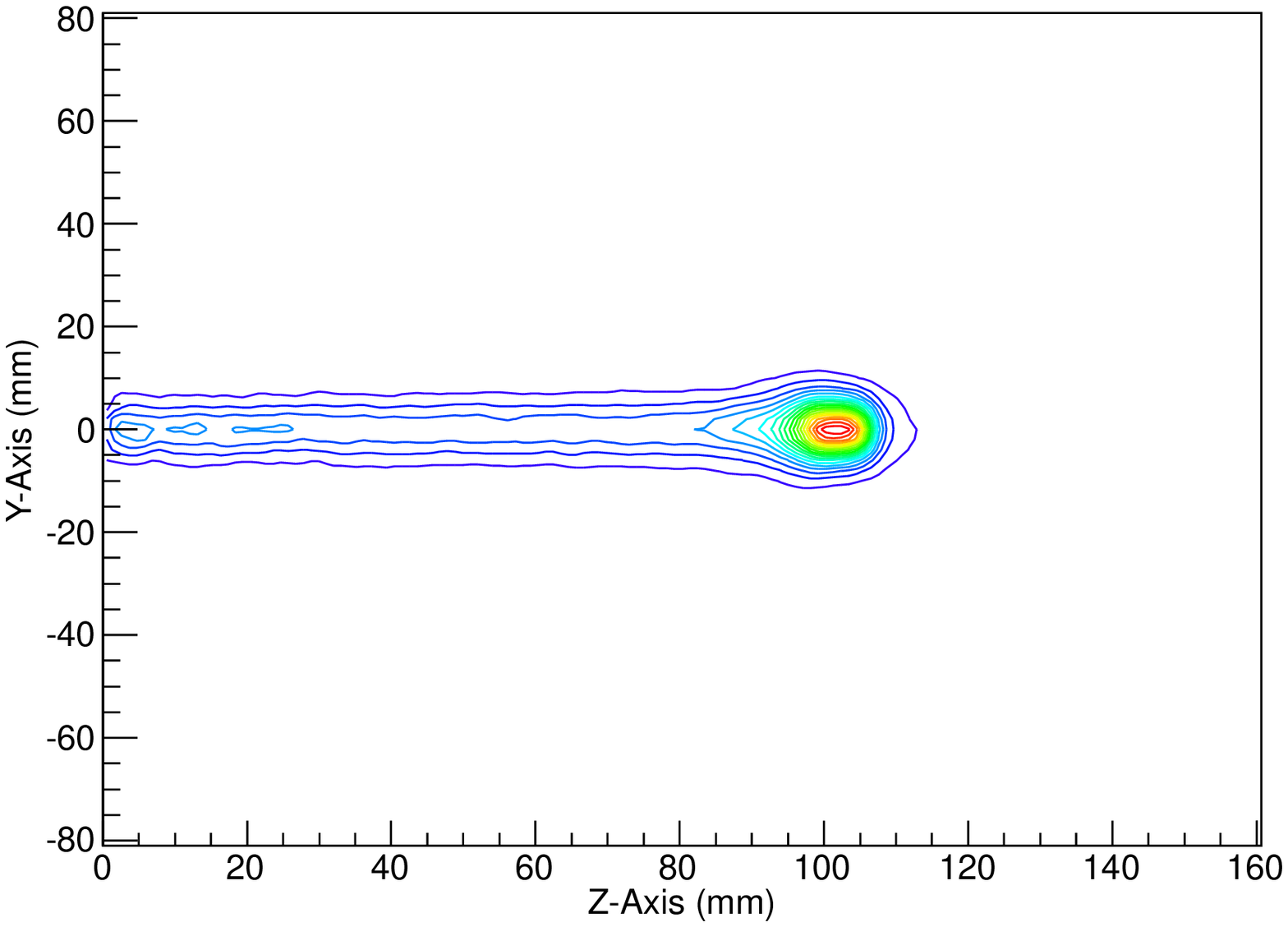}
    }
    \subfloat
    {
        \includegraphics[width=3.8cm]{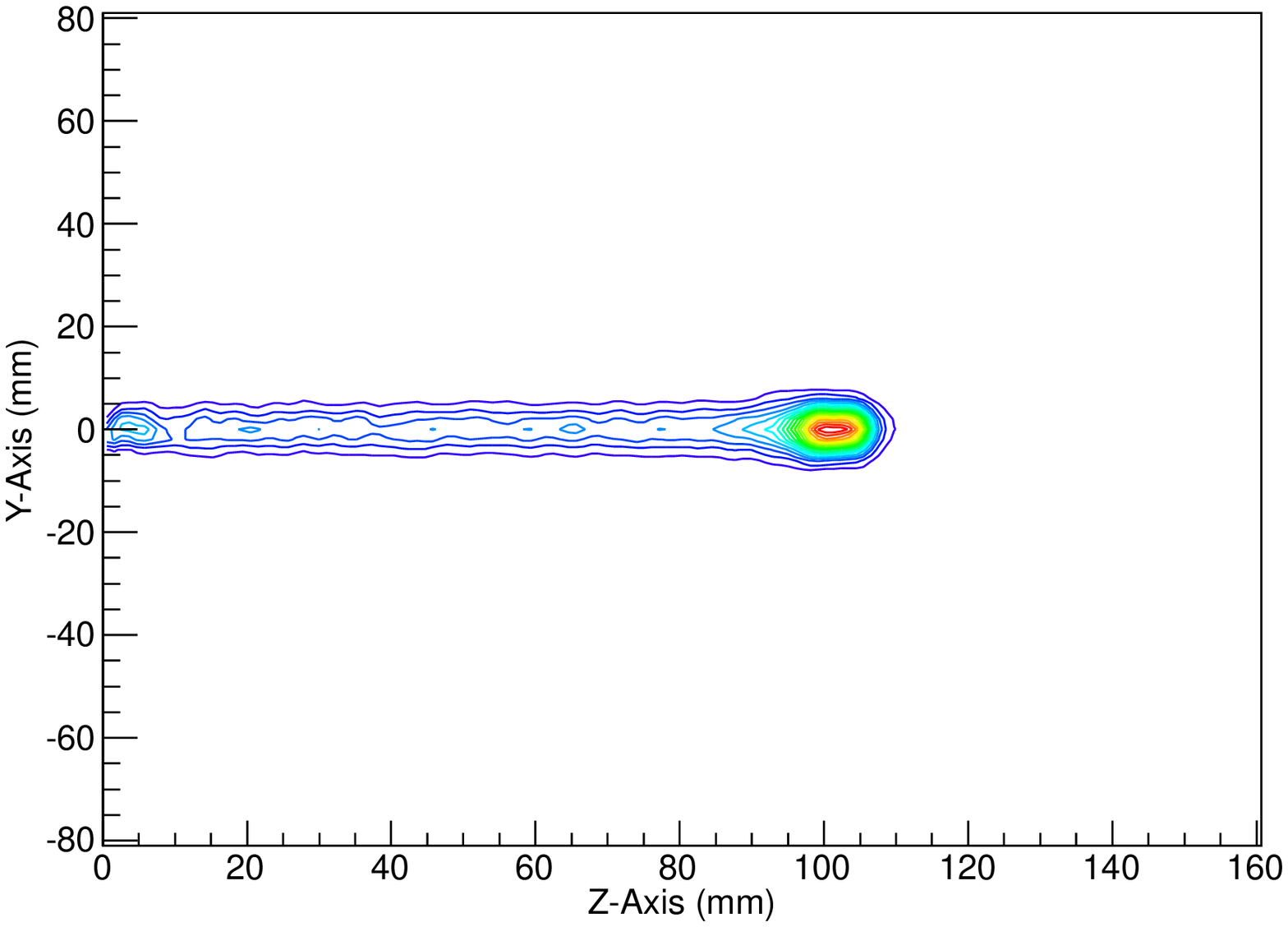}
    }
\\

    \subfloat[Dual-plate]
    {
        \includegraphics[width=3.8cm]{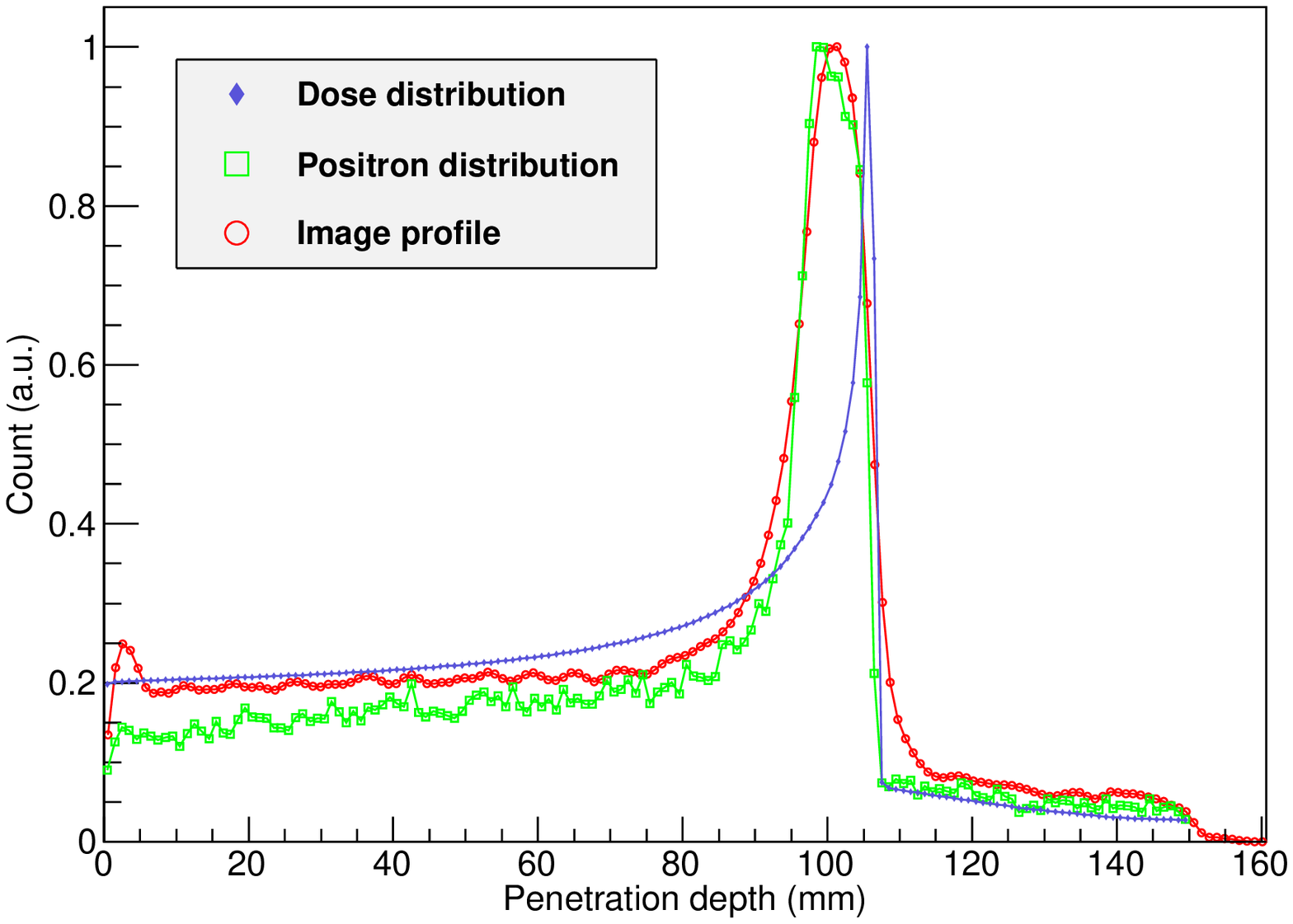}
    }
    \subfloat[Four-head]
    {
        \includegraphics[width=3.8cm]{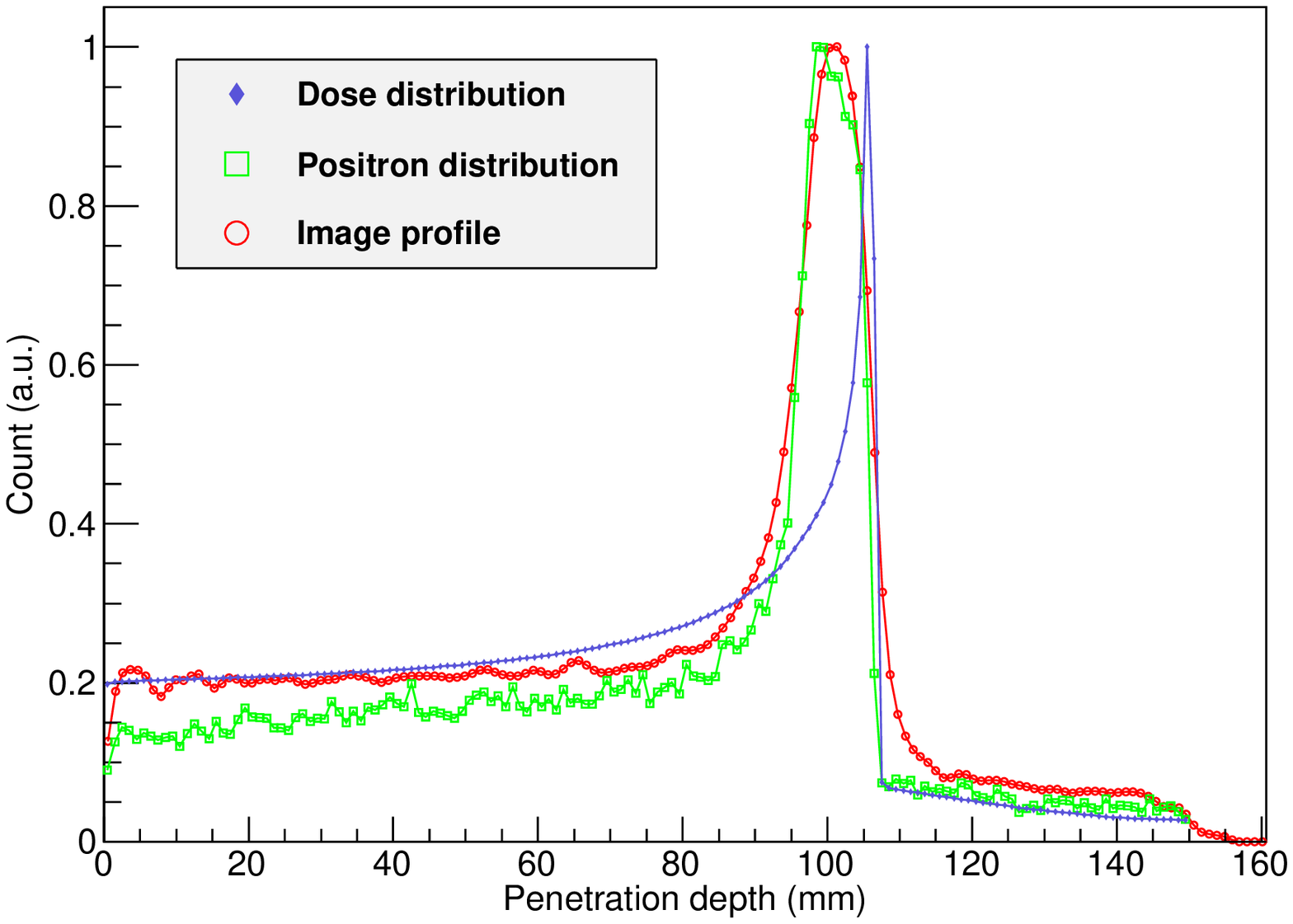}
    }
    \subfloat[Full-ring]
    {
        \includegraphics[width=3.8cm]{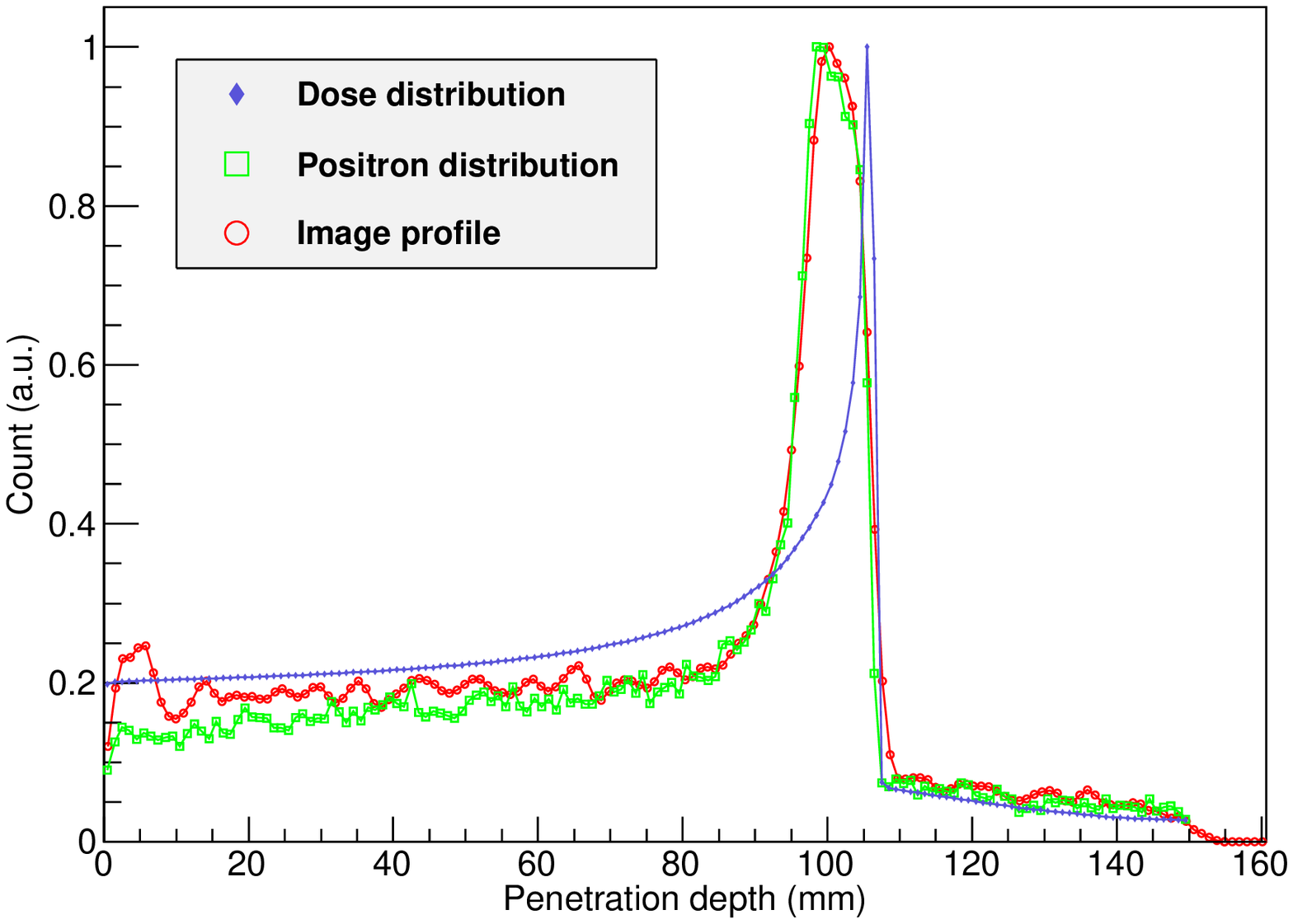}
    }
\end{center}

    \caption{Reconstructed PET images and their longitudinal profiles for (a) 172, (b) 200, (c) 250 AMeV $^{12}$C nuclei in the PMMA phantom. The simulated dose and $\beta^{+}$ activity distributions are also shown for comparison.}
    \label{fig_profile}
\end{figure}
\begin{multicols}{2}

Although the two kind of peaks are not overlapping together, there is a correlation of the dose and positron activity falloff at the distal edge~\cite{0031-9155-56-5-004}.
Parodi and Bortfeld~\cite{0031-9155-51-8-003} demonstrated a feasibility of dose reconvery from the positron distribution, which indicates the posibility of using PET to monitor the carbon beam therapy.

\subsection{PET images}
We use the 3D MLEM algorithm to reconstruct the activity distribution images for the three kinds PET scanner, and use the ROOT software to analyze the images.
In order to compare the performance of the scanners, the logitudinal profiles of reconstructed images, positron activity distribution and dose distribution are show together in Fig.\ref{fig_profile}.

The location of 172, 200 and 250 AMeV carbon beam was at left, center and right of the FOV, respectively.
In all these three cases, there was no significant difference between the image obtained by the dual-plate, four-head and full-ring scanners.
Consequently, the dual-plate PET scanner is feasible to monitor the dose distribution for carbon ion therapy.

\section{Conclusions and future work}
We proposed a dual-plate PET scanner for the dose verification in carbon ion therapy.
In the simulation study, the dual-plate scanner could avoid interference with the beam line and showed its feasibility to monitor the dose distribution.
Although the image performance of the dual-plate is worse than that of the four-head and full-ring scanner, especially at the periphery of the FOV, which is due to the planar nature of the data.

The gap between the Bragg peak and the profile of the reconstructed image is caused by the different physical porcesses that the dose distribution and the produced distribution of positron emitters undergo.
So, it is not possible to directly evaluate the dose distribution using the reconstructed image.
But, there does exist a correlation of the dose and positron activity falloff at the distal edge.
By using special fileter function for phantoms in one dimension within the area of the distal falloff of the dose, the dose distribution can be recoveried from positron acitivty distribution~\cite{0031-9155-51-8-003}.

With the dual-plate scanner, the maximum yeild position of positron emitters is successfully measured,
and the logitudinal profiles of reconstructed images, positron activity distribution match well.

In conclusion, the results of this simulaiton indicate that the dual-plate PET scanner is feasible to monitor the carbon ion therapy.

\end{multicols}

\vspace{-1mm}
\centerline{\rule{80mm}{0.1pt}}
\vspace{2mm}

\begin{multicols}{2}

\end{multicols}

\clearpage
\end{document}